\documentclass[aps,prb,twocolumn,showpacs,preprintnumbers,amsmath,amssymb,floatfix,superscriptaddress,10pt]{revtex4-1}
\usepackage{bm}
\usepackage{color}
\usepackage{float}
\usepackage{dcolumn}
\usepackage{graphicx}
\usepackage{booktabs}
\usepackage{hyperref}
\usepackage{multirow}
\usepackage{longtable}

\newcommand{\kfr}{{\bf k}_{\rm f}}
\newcommand{\kto}{{\bf k}_{\rm t}}
\newcommand{\exg}{\Delta_{\text{Ex}}}
\newcommand{\qpg}{\Delta_{\text{QP}}}

\newcolumntype{d}[1]{D{.}{.}{#1}}
\newcommand\mc[1]{\multicolumn{1}{c}{#1}}

\usepackage[dvipsnames]{xcolor}

\begin{document}

\title{Quantum Monte Carlo calculations of energy gaps from first principles}

\author{R.\ J.\ Hunt}
  \email{r.hunt4@lancaster.ac.uk}
\affiliation{Department of Physics, Lancaster University, Lancaster LA1 4YB,
United Kingdom}

\author{M.\ Szyniszewski}
\affiliation{Department of Physics, Lancaster University, Lancaster LA1 4YB,
United Kingdom}

\author{G.\ I.\ Prayogo}
\affiliation{Japan Advanced Institute of Science and Technology (JAIST), School
of Information Science, Asahidai 1-1, Nomi, Ishikawa 923-1292, Japan}

\author{R.\ Maezono}
\affiliation{Japan Advanced Institute of Science and Technology (JAIST), School
of Information Science, Asahidai 1-1, Nomi, Ishikawa 923-1292, Japan}

\affiliation{Computational Engineering Applications Unit, RIKEN,
2--1 Hirosawa, Wako, Saitama 351-0198, Japan}

\author{N.\ D.\ Drummond}
\affiliation{Department of Physics, Lancaster University, Lancaster LA1 4YB,
United Kingdom}

\date{\today}

\begin{abstract}
We review the use of continuum quantum Monte Carlo (QMC) methods for
the calculation of energy gaps from first principles, and present a
broad set of excited-state calculations carried out with the
variational and fixed-node diffusion QMC methods on atoms, molecules,
and solids. We propose a finite-size-error correction scheme for bulk
energy gaps calculated in finite cells subject to periodic boundary
conditions. We show that finite-size effects are qualitatively
different in two-dimensional materials, demonstrating the effect in a
QMC calculation of the band gap and exciton binding energy of monolayer
phosphorene. We investigate the fixed-node errors in diffusion Monte
Carlo gaps evaluated with Slater-Jastrow trial wave functions by
examining the effects of backflow transformations, and also by
considering the formation of restricted multideterminant expansions
for excited-state wave functions. For several molecules, we examine
the importance of structural relaxation in the excited state in
determining excited-state energies. We study the feasibility of using
variational Monte Carlo with backflow correlations to obtain accurate
excited-state energies at reduced computational cost, finding that
this approach can be valid. We find that diffusion Monte Carlo gap
calculations can be performed with much larger time steps than are
typically required to converge the total energy, at significantly
diminished computational expense, but that in order to alleviate
fixed-node errors in calculations on solids the inclusion of backflow
correlations is sometimes necessary.
\end{abstract}


%

\maketitle

\section{Introduction\label{sec:introduction}}

Accurate determination of the excited-state properties of atoms,
molecules, and solids is an outstanding goal of modern theoretical and
computational physics.  In the past few decades, progress has been
made in many avenues. Methods for calculating excitation energies from
first principles include density functional theory (DFT) and its
time-dependent extension, many-body perturbation theory, mainly in the
$GW$ approximation, the various quantum chemistry methods, e.g.,
configuration interaction and coupled-cluster methods, and also the
continuum quantum Monte Carlo (QMC) methods that we study here.

All of these methods have associated strengths and
weaknesses. Kohn-Sham DFT, while having reasonable computational cost
[$O(N^3)$ for a system of $N$ electrons], suffers the well-documented
band-gap problem,\cite{perdew1985density, seidl1996generalized}
whereby electronic band gaps are systematically underestimated. It has
been repeatedly shown that hybrid exchange-correlation functionals,
e.g., the B3LYP\cite{lee1988development,becke1993density} and HSE06
functionals,\cite{Heyd2003} which include a finite fraction of the
exact exchange energy, go some way towards remedying this problem, with
significant improvements being obtained for energy gaps in a range of
systems.\cite{heyd2005energy, muscat2001prediction,Paier2006} Newer
functionals incorporating screened exchange contributions have also
demonstrated improvements over standard DFT\@.\cite{clark2010}
Approaches based on many-body perturbation theory in the $GW$
approximation have proven to be very effective in determining
the excited-state properties of weakly-correlated
solids.\cite{Hybertsen1985,Hybertsen1986,Godby1986,Godby1988} However,
$GW$ results obtained under different levels of self-consistency can
often disagree substantially, and $GW$ results themselves can depend
significantly on underlying single-particle orbital-generation
calculations (i.e., on the particular $G_0$ and $W_0$ used to enter
the self-consistent cycle).\cite{bruneval2012benchmarking,Blase2011}
The coupled-cluster and configuration-interaction methods, although
very accurate in the description of small systems, scale very poorly
with system size. The computational cost of most coupled-cluster
implementations scales as $O(N^p)$, where $p$ is a relatively high
power (e.g., $p=7$ for coupled cluster including single and double
excitations, with triples treated perturbatively), and configuration
interaction scales exponentially with $N$. This renders any
application of these methods to solids prohibitively expensive. Full
configuration interaction QMC is another example of a highly accurate
method which has recently been used to study excited states, but
ultimately shows the same exponential scaling as configuration
interaction, albeit with a significantly smaller
prefactor.\cite{booth2012communication,blunt2015excited}

Continuum QMC techniques,\cite{Foulkes1999,ceperley1986quantum} on the
other hand, offer us an accurate means of probing both ground- and
excited-state properties of atoms, molecules, and solids from first
principles and with excellent system-size scaling. Without backflow
correlations,\cite{lopezrios2006inhomogeneous,kwon1998effects} the
computational cost of QMC scales as $O(N^3)$, as with DFT, although
the prefactor is typically over a thousand times larger. With
backflow, the cost scaling incurs an additional factor of $N$. QMC has
previously been used to study the excited-state properties of
silicon,\cite{Kent1998,Williamson1998}
diamond,\cite{Towler2000,mitasbook}
hydrogenated silicon clusters,\cite{porter_2001_excitons}
diamondoids,\cite{drummond2005electron}
solid hydrogen,\cite{azadi2017}
solid nitrogen,\cite{mitas1994quantum}
zinc oxide and selenide,\cite{yu2015}
vanadium dioxide,\cite{Zheng2015}
nickel oxide,\cite{Mitra2015}
manganese nickelate,\cite{Dzubak2017}
the two-dimensional (2D) homogeneous electron gas
(HEG),\cite{kwon1994,holzmann2009,drummond2013diff,drummond2013}
Rydberg states,\cite{bande_2006}
and various molecular
systems.\cite{grimes1986,reynolds1986,bernu1990,williamson2002,aspuru2004,
mostaani2016,porter2001,grossman_2001,schautz_2004,tiago2008}

In variational Monte Carlo (VMC), expectation values of observables
with respect to explicitly correlated trial wave functions are
evaluated using Monte Carlo integration techniques. Starting with a
product of Slater determinants of single-particle orbitals
$\{\phi^{\uparrow}\}$ and $\{\phi^{\downarrow}\}$ for up-spin and
down-spin electrons,
\begin{equation}
\mathcal{D}({\textbf R}) = \det{[\phi^{\uparrow}_{i}({\bf r}_j)]}
\det{[\phi^{\downarrow}_{k}({\bf r}_l)]}, \label{eq:swfn}
\end{equation}
with ${\bf R} = \left( {\textbf r}_1, {\textbf r}_2, \ldots, {\textbf r}_N
\right)$, $i,j \in \{1,\ldots,N_{\uparrow} \}$, and $k,l \in
\{N_{\uparrow}+1,\ldots,N\}$, we form the so-called
Slater-Jastrow\cite{Foulkes1999} (SJ) trial wave function
\begin{equation}
  \Psi_{{\rm SJ}}({\textbf R})= \exp{\left[ J({\textbf R}) \right]}
  \cdot\mathcal{D}({\bf R}),\label{eq:sjwfn}
\end{equation}
where $J({\textbf R})$ is the Jastrow exponent, and
$\exp\left[J\right] > 0$.  The Jastrow factor is an explicit function
of interparticle coordinates containing optimizable parameters, and
allows the many-electron trial wave function to obey the Kato cusp
conditions.\cite{kato1957eigenfunctions} However, since $\exp(J)>0$,
the Jastrow factor does not affect the nodal surface of the trial wave
function.

We have also made use of Slater-Jastrow-backflow (SJB) wave functions,
to improve the nodal surfaces of our wave functions. The backflow
transformation corresponds to the replacement ${\bf R} \rightarrow
{\bf X}({\bf R})$ in Eq.\ (\ref{eq:swfn}), with ${\bf X} = \left({\bf x}_1,
{\bf x}_2,\ldots,{\bf x}_N\right)$ being the ``collective'' or
``quasiparticle'' coordinates. Each of these new coordinate vectors
${\bf x}_i({\bf R})$ depends on all the particle positions and is given by
\begin{equation}
  {\bf x}_i = {\bf r}_i + \boldsymbol\xi_i({\bf R}),
\end{equation}
where ${\boldsymbol\xi}$ is the backflow displacement. The resulting
many-body trial wave function is labeled $\Psi_{{\rm SJB}}$, and in
general has a nodal surface that differs from $\Psi_{{\rm SJ}}$ when
evaluated with the same single-particle orbitals and Jastrow factor.
Provided the backflow displacement ${\boldsymbol\xi}$ is a smooth
function of ${\bf R}$, backflow describes a smooth transformation of
space under the Slater wave function, and is not therefore expected to
alter the nodal surface qualitatively (i.e.\ backflow cannot create nor
destroy individual nodal pockets).  Hence backflow does not
address the issue of static correlation; however, in the context of
excited-state calculations the fact that backflow does not alter nodal
topology is useful, as it ensures that the SJB trial wave function
describes the same state of the system as the SJ wave function.
These trial wave functions describe excited states of the
interacting system that are adiabatically connected to excited states of the
noninteracting system. The topology
of the nodal surface, and its bearing on excited state QMC calculations, is
further discussed in Sec.\ \ref{sub:nodal_topology}.  Multideterminant
wave functions, which can change nodal topology, are discussed in Sec.\
\ref{subsub:mdets}.

In diffusion Monte Carlo (DMC), a wave function $\phi_{\rm DMC}$ is
evolved in imaginary time with the use of stochastic techniques, such
that each excited-state component decays exponentially with imaginary
time at a rate proportional to its total energy. In fixed-node DMC,
the nodal surface is fixed to that of the trial wave function; the set
of points for which $\phi_{\text{DMC}} = 0$ (the DMC nodal surface)
coincides with the set of points for which $\Psi_{\text{SJ(B)}} = 0$
(the trial nodal surface). This means that the DMC algorithm projects
out and samples the lowest-energy state that is compatible with a
given trial nodal surface. This leads to the well-known fixed-node
error, which prevents the numerically exact evaluation of many-fermion
ground-state total energies in polynomial time.  However, the
fixed-node approximation is the only tractable way in which we are
able to calculate excited-state energies in DMC\@. By forming trial
excited states, and fixing their nodes, we can evaluate excited-state
energies. If the nodal surface of a trial excited-state wave function
is exact, the DMC energy of that excited state is also exact.

In this article, we present the results of a systematic study of
static-nucleus energy gaps for various atoms, molecules, and solids
obtained using the VMC and DMC methods.  The rest of the article
proceeds as follows. In Sec.\ \ref{sec:excited_qmc} we present the
theoretical background on the QMC evaluation of energy gaps, the
treatment of finite-size effects, and other technical aspects of
excited-state QMC calculations.  In Sec.\ \ref{sec:methodology} we
discuss the computational details of our example calculations, the
results of which are presented in Sec.\ \ref{sec:results}. Finally,
our conclusions are drawn in Sec.\ \ref{sec:conclusions}.  Hartree
atomic units ($\lvert e\rvert =m_{\rm e}=4\pi\epsilon_0=\hbar=1$) are used
throughout, unless otherwise stated.

\section{Excited-state QMC\label{sec:excited_qmc}}

\subsection{Quasiparticle and excitonic gaps}\label{sub:qp_ex_gaps}

In order to perform a QMC supercell calculation with periodic boundary
conditions, the trial wave function must satisfy the many-body Bloch
conditions outlined in Ref.\ \onlinecite{Rajagopal1995}.
Specifically, the wave function should acquire a phase $\exp(i{\bf
  k}_{\rm s}\cdot{\bf R}_{\rm s})$ whenever a single particle is
translated through a supercell lattice point ${\bf R}_{\rm s}$, where
the constant vector ${\bf k}_{\rm s}$ is the \textit{supercell Bloch
  vector} or \textit{twist}.  Furthermore, the wave function should
acquire a phase $\exp(i{\bf k}_{\rm p}\cdot {\bf R}_{\rm p})$ when all
the particles are together translated through a primitive lattice
point ${\bf R}_{\rm p}$, where ${\bf k}_{\rm p}$ lies in the first
Brillouin zone of the primitive cell. This is usually achieved by
requiring the Jastrow factor and backflow function to have the
periodicity of the supercell under single-particle displacements and the
periodicity of the primitive cell under all-particle displacements, while
the Bloch orbitals in the Slater
determinant lie on a regular grid of primitive-cell reciprocal lattice
points offset by the supercell Bloch vector ${\bf k}_{\rm s}$.  E.g., for an
$l\times m\times n$ supercell, this grid would be an $l\times m\times n$ grid
of ${\bf k}$-points in the primitive-cell Brillouin zone), centered on
the supercell Bloch vector ${\bf k}_{\rm s}$. Folding of these points
into the supercell Brillouin zone results in all points being
mapped to ${\bf k}_{\text{s}}$. The occupancies of the single-particle
orbitals at each ${\bf k}$ point in the primitive-cell Brillouin zone
can then be used define excitations.

The \textit{quasiparticle gap} $\qpg$ of a system is the energy
required to create an unbound electron-hole pair in that system. It is
given by the difference between a conduction-band minimum ${\cal
  E}_{\rm CBM}$ and a valence-band maximum ${\cal E}_{\rm
  VBM}$,\footnote{In a finite system, the ``conduction-band minimum''
  is $-A$, where $A$ is the electron affinity and the ``valence-band
  maximum'' is $-I$, where $I$ is the first ionization potential.}
i.e.,
\begin{align}
  &\qpg(\kfr,\kto)={\cal E}_{\text{CBM}}(\kto)-{\cal E}_{\text{VBM}}(\kfr)
  \nonumber \\
  &=\left[E_{N+1}(\kto)-E_N(\kto)\right]-
    \left[E_N(\kfr)-E_{N-1}(\kfr)\right] \nonumber \\
  &=E_{N+1}(\kto)+E_{N-1}(\kfr)-E_N(\kto)-E_N(\kfr),
\end{align}
where $E_N$ is the total ground-state energy of an $N$-electron
system. The labels $\kfr$ and $\kto$ denote the ${\bf k}$-points from
which and to which excitations are made, and may
be ignored in finite systems. The ground-state energies $E_N(\kto)$
and $E_N(\kfr)$ are identical if the calculations used to evaluate the
quasiparticle energies $E_{N\pm1}$ are performed on the same grid of
${\bf k}$-vectors [i.e., for cells with the same supercell Bloch
vector ${\bf k}_{\text{s}}$ we have
$\qpg(\kfr,\kto)=E_{N+1}(\kto)+E_{N-1}(\kfr)-2E_N$]; otherwise, they
may differ. It is always possible to evaluate $\qpg$ between any pair
of ${\bf k}$-points $\kfr$ and $\kto$ at any system size by
appropriate choices of the supercell Bloch vector ${\bf k}_{\text{s}}$
(i.e., the offset of the ${\bf k}$-point grid) in the two cases.

The \textit{excitonic gap} (or \textit{optical gap}) of a system is
the energy required to create a bound electron-hole pair in that
system. It is given by the difference of total energies obtained with
an electron promoted to an excited state of the system and the total
energy of the ground state
\begin{equation}
  \exg(\kfr,\kto) = E^{+}_N(\kfr, \kto) - E_N,
\end{equation}
with $E^{+}_N(\kfr, \kto)$ the \textit{excited-state} total energy
of an $N$-electron system in which an electron has been promoted from an
occupied valence-band orbital at $\kfr$ to an unoccupied conduction-band
orbital at $\kto$ (again, the ${\bf k}$-point labels may be ignored in the
finite case). The ground-state energy $E_N$ is in this case unambiguous, and
has to be evaluated with the same ${\bf k}$-point grid as the excited-state
energy $E^{+}_N(\kfr,\kto)$. In the rest of this section, we will suppress the
${\bf k}$-point labels $\kfr$ and $\kto$. Note that, unlike the quasiparticle
gap, the excitonic gap may only be evaluated between pairs of ${\bf k}$-points
that are \textit{simultaneously} included in the ${\bf k}$-point grid (i.e.,
the set of ${\bf k}$ points must contain both $\kfr$ and $\kto$). This is not
generally possible for a given pair of ${\bf k}$-points at all system sizes.
For example, it is possible to calculate a vertical excitonic gap ($\kfr=\kto$)
in any supercell by using an appropriate offset ${\bf k}_{\rm s}$ to the grid
of ${\bf k}$ vectors; however, it is only possible to calculate an excitonic
gap from $\Gamma$ to K in a 2D hexagonal cell in supercells of $3l \times 3m$
primitive cells, where $l$ and $m$ are integers.

For our purposes, the total energy $E_{N-1}$ ($E_{N+1}$) is evaluated
by calculation of the QMC energy of a state with the removal
(addition) of an electron from (into) an occupied (unoccupied) state
in the Slater determinant. Similarly, the total energy $E^{+}_N$ is
evaluated by calculation of the QMC energy of a state whose valence-
and conduction-band occupancies have been switched for the particular
orbitals of interest. This trial wave function describes a correlated state
of an excited electron and remnant hole, i.e.,
an exciton. The difference $E^{\rm X}_{\rm B}=\qpg-\exg$ is equal to
the exciton binding energy for a particular configuration of electron
and hole, and is always greater than or equal to zero for a finite
system or for an extended system in the thermodynamic limit, because
the electron-hole Coulomb interaction is attractive. This may not be
the case in QMC data obtained in a finite periodic cell, in which case
finite-size effects may lead to the apparently unphysical scenario
where $\qpg<\exg$. The origin of this behavior is explained in
Sec.\ \ref{sub:fs_effects}. The exciton binding energy $E_{\rm B}^{\rm
  X}$ can only be evaluated at system sizes for which calculation of
$\exg$ is permitted. It may be reexpressed as
\begin{align}
  &E^{\rm X}_{\rm B}(\kfr,\kto)=
  \qpg(\kfr,\kto)-\exg(\kfr, \kto) \nonumber \\
  &= E_{N+1}(\kto)+E_{N-1}(\kfr) - E^{+}_{N}(\kfr, \kto)-E_N,
\label{eq:ex_binding}
\end{align}
with $E_N=E_N(\kfr)=E_N(\kto)$. The four QMC total energies in
Eq.\ (\ref{eq:ex_binding}) are statistically independent, unlike
$\qpg$ and $\exg$, which both depend on the same ground-state energy
$E_N$.

\subsection{Singlet and triplet excitations}\label{sub:sing_trip}

In the preceding section, we neglected to include information on the possible
spin degree of freedom of the electrons involved in excitations.  For
Hamiltonians that include no spin-orbit coupling we can, with no added
difficulty, define the quasiparticle and excitonic gaps including explicitly
the spin $\sigma \in \{\uparrow,\downarrow\}$ of the electron, which is excited
from $(\kfr$, $\sigma_{\rm f})$ to $(\kto$, $\sigma_{\rm t})$. Singlet
excitations are those with $\sigma_{\rm f}=\sigma_{\rm t}$, while triplet
excitations incur a spin flip, $\sigma_{\rm f} \neq \sigma_{\rm t}$. In QMC,
the spin of any electrons involved in excitations can be controlled by
specification of the (spin-dependent) orbital occupancies in the Slater part of
the trial wave function. In most cases singlet excitations are more physically
relevant, because triplet optical excitations are forbidden in first-order
perturbation theory.  The feasibility of calculating singlet-triplet splittings
by QMC techniques depends on the magnitude of the singlet-triplet splitting;
the resolution of a small energy difference requires small QMC statistical
error bars. We have calculated the singlet-triplet splitting of the lowest
lying excitonic states of anthracene in Sec.\ \ref{subsub:non_dimers} and of
the ground state of O$_2$ in Sec.\ \ref{subsub:o2_dimer}.

\subsection{Wave-function nodes and variational
principles}\label{sub:wfn_nodes_vps}

DMC gives the energy of any eigenstate of the Hamiltonian exactly if
the nodal surface of the trial wave function is
equal to that of the true eigenfunction, even if the trial wave
function is approximate between the nodes.\cite{Foulkes2001} In
general, however,
each of the total energies $E_N$, $E_{N\pm1}$, and $E^+_N$ suffers a
fixed-node error due to the inexact nodal surface of the trial wave
function. Assuming the excitations are made into the lowest-energy
quasiparticle bands, $E_N$ and $E_{N\pm1}$ are themselves ground-state
total energies, and hence the fixed-node errors in $E_N$ and $E_{N\pm1}$ must
be positive.\cite{Foulkes2001} There is no rigorous variational principle on
the quasiparticle gap $\qpg=E_{N+1}+E_{N-1}-2E_N$, although in practice gaps
evaluated using total
energies evaluated by a variational method usually provide upper bounds. In
Hartree-Fock theory, the absence of electronic correlation has the consequence
that electrons localize excessively to avoid one another, and hence
quasiparticle energy gaps are overestimated significantly. For example, in Si
the Hartree-Fock quasiparticle gap is an overestimate by around 4.5
eV\@.\cite{Dovesi1981,Ohkoshi1985} In QMC, as we recover more and more of the
electronic correlation energy by optimizing Jastrow factors and backflow
functions, and performing DMC to project out the fixed-node ground state, we
observe that quasiparticle gaps reduce substantially from their Hartree-Fock
values towards their exact static-nucleus nonrelativistic values.  Apart from
the unlikely case in which we recover significantly more correlation energy in
the $(N\pm1)$-electron systems compared to the ground-state $N$-electron
system, we therefore expect QMC quasiparticle gaps to be upper bounds on the
exact gaps. Because individual contributions to the quasiparticle gap
separately obey ground-state variational principles, one expects to obtain
improved DMC estimates of quasiparticle gaps by reoptimizing parameters that
affect the nodal surfaces in the $(N\pm 1)$-electron systems.  Improving a
Jastrow factor is expected to improve VMC energy gaps, but not fixed-node DMC
gaps, since the Jastrow factor does not affect the nodal surface.
(Of course, improving the Jastrow factor reduces statistical error
bars, finite-time-step bias, finite-population bias, and
pseudopotential locality errors; furthermore, parameters that do affect
the nodal surface should be optimized together with the Jastrow
factor.)

Let us now consider the fixed-node error in the excitonic gap.  Again,
the ground-state energy can only be overestimated by the fixed-node
variational principle. The excited-state energy $E^{+}_N$, however, is
not bounded by variational principles except in special
circumstances. If the trial excited-state wave function transforms
as a one-dimensional (1D) irreducible representation (irrep) of the full
symmetry group of the many-body Hamiltonian, then the resultant
fixed-node DMC energy provides an upper bound on the energy of the
lowest-lying eigenstate that transforms as that 1D
irrep.  In that case, the error in the DMC energy
is second order in the error in the nodal surface of the
excited-state trial wave function,
and there is a tendency for positive fixed-node errors to cancel in
excitonic gaps. In the likely case that we recover more correlation
energy in the ground state than in the excited-state calculation,
QMC excitonic gaps act as upper bounds to their exact
counterparts.

If, however, the trial excited-state wave function does not transform
as a 1D irrep, or we are not
studying the lowest-energy eigenstate that transforms as the same
irrep as the trial wave function then the
fixed-node error in the excited-state energy $E^{+}_N$ can be either
positive or negative, and hence there could be cases in which the DMC
excitonic gap is too small. As a consequence, reoptimization of
trial-wave-function parameters affecting the nodal surface can lead to
absurd results, as the nodal surface becomes more like that of the
ground state. We provide an example illustrating this behavior in
Sec.\ \ref{sub:atoms}.

If the excited-state trial wave function transforms as a
multidimensional irrep of the full symmetry group
of the Hamiltonian, then weaker lower bounds on the estimate of the
excited-state energy can be realized
by forming trial wave functions that transform as 1D
irreps of subgroups of the full symmetry group of
the Hamiltonian.\cite{Foulkes1999} This is discussed in
Sec.\ \ref{subsub:mdets}.

Importantly, for excitations made between different ${\bf k}$ points, where
complex Bloch states (having definite crystal momentum ${\bf k}_{\text{T}}$)
can be chosen to populate the Slater part of the trial wave function,
variational principles on the lowest energy excitations are always realized
because of translational invariance (states of definite crystal momentum
transform according to 1D irreps of the space group, in line with the many-body
Bloch conditions).\cite{Foulkes1999} In the case where one wishes to form real
linear combinations of complex Bloch states with crystal momenta ${\bf
k}_{\text{T}}$ and ${-\bf k}_{\text{T}}$, respectively, the subsequent real
superposition does not generally transform as a 1D irrep of the space group,
and hence excited-state variational principles are not in general realized. If
${\bf k}_{\text{T}}$ happens to be on the edge of the Brillouin zone, however,
${\bf k}_{\text{T}}$ and ${-\bf k}_{\text{T}}$ are equivalent, and an
excited-state variational principle is realized once again. If one is not able
to recover an excited-state variational principle in this way, then one should
use complex Bloch orbitals (maintaining a variational principle, at the cost of
added computational expense). The so-called \textit{fixed-ray} method of Hipes
has been developed specifically to ensure the existence of excited-state
variational principles in cases of degeneracy such as this.\cite{Hipes2011}

Variational bounds on excited-state energies may also be obtained by
other means, e.g.\ via MacDonald's theorem.\cite{MacDonald1933} Zhao
and Neuscamman have recently devised a method which allows for the
realization of a variational principle on selected excited-state
energies, and also for practical optimization of excited-state QMC
trial wave functions.\cite{Zhao2016} Mussard \textit{et al.}\ have
extended the VMC method using the ideas of time-dependent
linear-response theory to extract excited-state properties, and have
presented example calculations within the Tamm-Dancoff approximation
to the linear-response equations.\cite{Mussard2018}

\subsection{Nodal topology}\label{sub:nodal_topology}

Fixed-node DMC works by obtaining exact ground-state solutions to the
Schr\"{o}dinger equation within nodal pockets, i.e., within the
regions of configuration space bounded by the nodes of the trial wave
function.\cite{Foulkes2001} The boundary conditions on the
Schr\"{o}dinger equation in each nodal pocket are that the DMC wave
function goes to zero at the edges of the pocket.  If the nodes of an
excited-state trial wave function are exact then the ground-state
energy in each nodal pocket is equal to the excited-state energy
corresponding to the trial wave function.

From the point of view of fixed-node DMC, the fundamental differences between
the ground-state many-electron wave function and its excited-state counterparts
are codified in the topology of their respective nodal surfaces, which
completely determine the corresponding fixed-node DMC energies.  The nodal
surface of the many-electron ground state satisfies a tiling property (all
nodal pockets are equivalent under permutations of identical fermions; this
is also true of determinants of Kohn-Sham orbitals),\cite{Ceperley1991} and it
is conjectured that the presence of only two nodal pockets is a generic feature
of the many-electron ground state.\cite{Mitas2006,Bressanini2012}
The nodal surfaces of excited states are
less-well-understood; they do not satisfy a tiling property in general unless
the trial state transforms as a 1D irrep
of the group of the Hamiltonian, and in
the general case the number of nodal pockets they possess can only be bounded:
Hilbert and Courant\cite{CourantMethods} proved that the nodes of the
$n^{\text{th}}$ excited state
divide configuration space into no more than $n+1$ nodal pockets.
\footnote{In 1D, a rigorous analysis of the topology of the ground and
  excited-state nodal surfaces culminates in the Hilbert-Courant nodal
  line theorem. The ground state is nodeless, and the $n^{\text{th}}$
  (non-degenerate) excited-state has $n$ nodes, dividing the 1D
  configuration space into $n+1$ nodal pockets (saturating the earlier
  stated constraint, which applies in dimensions greater than one).}
The fact that the number of inequivalent nodal pockets remains small
in low-lying excited states means that, for a sufficiently large DMC
target population, each set of equivalent nodal pockets will have a
significant initial population of walkers; furthermore, the walker
populations in high-energy sets of pockets are expected to die out on
an imaginary-time scale given by the inverse of the difference between
the energies of the different nodal pockets.  Hence the fixed-node DMC
energy with an excited-state trial wave function is equal to the
lowest of the pocket ground-state energies.  An example of this
behavior is shown in Sec.\ \ref{subsub:h_nodes}.

It is not possible for a Jastrow factor to alter the nodal surface of a trial
state, and nor is it possible for a smooth backflow function to alter the
\textit{topology} of a trial state. It is this fact that prevents variational
collapse of excited-state energies in VMC calculations in the cases of
electron addition, removal, or promotion where the trial state is a state of
definite symmetry transforming as a 1D irrep. While nodal topology is an
important factor in the description of excited states in QMC, and it is
important that backflow functions preserve it, as we show in Sec.\
\ref{subsub:h_nodes}, the correct nodal topology does not guarantee that one
will obtain reasonable results when optimizing backflow functions in trial
excited states. In cases where trial wave functions do not transform as 1D
irreps of the symmetry group of the Hamiltonian, preserving the nodal
topology can
still lead to the formation of a pathological nodal surface and a DMC
energy which is too low.

We note that, while we will not explicitly consider their use here, pfaffian
and geminal pairing wave functions have recently been shown to be somewhat more
efficient at accurately describing the nodes of a few systems where the exact
nodes are known.\cite{Bajdich2008}

\subsection{Finite-size effects}\label{sub:fs_effects}

A major source of error in gap calculations for condensed matter using
explicitly correlated wave-function methods such as QMC is the
presence of finite-size (FS) effects. For calculations on solids, we
are only able to simulate a finite supercell subject to periodic
boundary conditions. This means that our raw DMC data contain unwanted
contributions from the electrostatic interaction of added (or removed)
charges with their periodic images, and we must either correct for
this effect or extrapolate to infinite supercell size. A
general simulation supercell in $d$ dimensions is defined by a $d\times d$
integer ``supercell matrix'' $S$, which expresses the supercell lattice
vectors $\{{\bf a}_i^{\rm sc}\}$ in terms of primitive-cell lattice vectors
$\{{\bf a}_i^{\rm prim}\}$:
\begin{equation}
{\bf a}_i^{\rm sc} = \sum_j S_{ij} {\bf a}_j^{\rm prim}.
\end{equation}
A ``diagonal supercell'' is one for which the supercell matrix is
diagonal; such a supercell consists of an $S_{11} \times S_{22} \times
S_{33}$ array of primitive cells. In general a supercell contains
$\det{(S)}$ primitive cells.

Various FS correction schemes exist for the total energies per primitive cell of
solids calculated at fixed system size in
DMC\@.\cite{Chiesa2006,drummond2008finite,Kwee2008} However, such FS errors
cancel between ground and excited states and are of little relevance to the FS
effects in excitation energies. Let us first consider the FS effects in $\qpg$.
The leading-order FS error is due to the self-interaction of added
quasielectrons or quasiholes. The energy of the resulting unwanted lattice of
quasiparticles (each having charge $q=\pm1$) is given by a screened Madelung
sum over supercells, i.e., $q^2 v_{\rm M}({\bf a}_1^{\rm sc},{\bf a}_2^{\rm
sc},{\bf a}_3^{\rm sc}) / 2$ with $v_{\rm M}({\bf a}_1^{\rm sc},{\bf a}_2^{\rm
sc},{\bf a}_3^{\rm sc})$ being the screened Madelung constant for the
supercell.\cite{Madelung1918} There are two separate terms of this type
in a quasiparticle gap correction, one for $-{\cal
E}_{\text{VBM}}=E_{N-1}-E_{N}$ and another for ${\cal
E}_{\text{CBM}}=E_{N+1}-E_N$.  A physically reasonable FS correction formula
for $\qpg$ therefore reads \begin{equation} \qpg(\infty) \approx \qpg({\bf
a}_1^{\rm sc},{\bf a}_2^{\rm sc},{\bf a}_3^{\rm sc}) - v_{\rm M}({\bf
a}_1^{\rm sc},{\bf a}_2^{\rm sc},{\bf a}_3^{\rm sc})\label{eq:qpg_corr},
\end{equation} where $\qpg(\infty)$ is the infinite-system quasiparticle gap.
A similar expression has previously been used at the DFT level to study FS
effects in the formation energies of charged defects.\cite{Hine2009}
Assuming the separation of the neighboring images of the quasiparticle is sufficiently
large that linear response theory is valid, $v_{\rm M}({\bf a}_1^{\rm
sc},{\bf a}_2^{\rm sc},{\bf a}_3^{\rm sc})$ can be evaluated using an
appropriately screened Coulomb interaction.  In QMC calculations with fixed
ions, only the electronic contribution to the susceptibility is relevant to
the FS effects in the quasiparticle gap, i.e., the permittivity that
should be used to evaluate the screened Madelung constant is the
high-frequency permittivity.  This can usually be evaluated with sufficient
accuracy using density functional perturbation theory,\cite{baroni2001} if
experimental results are unavailable.  In anisotropic materials, the Madelung
constant must be evaluated using the permittivity tensor, as is done in DFT
studies of charged defect formation energies.\cite{murphy2013} A simple
expression for the anisotropically screened Madelung constant can be obtained
by a coordinate transformation to the principal axes of the
permittivity tensor. If $\tilde{v}_{\rm M}({\bf a}_1^{\rm sc},{\bf a}_2^{\rm
sc},{\bf a}_3^{\rm sc})$ is the unscreened Madelung constant then the
screened Madelung constant is
\begin{eqnarray}
  & & v_{\rm M}({\bf a}_1^{\rm sc},{\bf a}_2^{\rm sc},{\bf a}_3^{\rm sc})
  \nonumber \\ & & \hspace{2em} {} = \dfrac{1}
  {\sqrt{\det{(\epsilon)}}}\tilde{v}_{\rm
  M}(\epsilon^{-1/2}{\bf a}_1^{\rm sc},\epsilon^{-1/2}{\bf a}_2^{\rm sc},
  \epsilon^{-1/2}{\bf a}_3^{\rm sc}), \label{eq:screened_vM}
\end{eqnarray}
where $\epsilon$ is the high-frequency permittivity tensor of the system. The
properties of physical
permittivity tensors mean that the square root of the inverse is always
well-defined: positive-definite matrices have only one square root, also known
as the principal square root.  This expression can be obtained from an analysis
of the Ewald interaction in the presence of an anisotropic medium (supplied in
the present case by the rest of the system). Similar arguments were given by
Fischerauer for the interaction between aperiodic point charges in
anisotropic media.\cite{fisch_greens_function} In the case of an isotropic
medium, Eq.\ (\ref{eq:screened_vM}) reduces to division of the unscreened
Madelung constant by the relative permittivity, i.e.,
$v_{\rm M}=\tilde{v}_{\rm M}/\epsilon$.

In layered and 2D materials, the in-plane polarizability of the layers
modifies the form of the Coulomb interaction to the so-called Keldysh
interaction.\cite{Rytova1965coulomb,Rytova1967screened,Keldysh1979}
Depending on the in-plane susceptibility and the spatial extent of the
simulation cell, it may be necessary to employ this modified form of
interaction in the evaluation of the screened ``Madelung''
constant. For supercells much larger than the length scale $r_*$
defined by the ratio of the in-plane susceptibility to the
permittivity of the surrounding medium, the Keldysh interaction
between image charges reduces to Coulomb form, and the subtraction of
the screened Coulomb Madelung constant is reasonable. On the other
hand, if the supercell size is significantly less than $r_*$ then the
Keldysh interaction is of logarithmic form\cite{Mostaani2017} and the
resulting Madelung constant is roughly independent of system size,
until the linear size of the simulation cell reaches $r_*$.  We
discuss this further in Sec.\ \ref{subsub:phosphorene}.

If the leading-order FS error in the quasiparticle gap is removed by
subtracting the screened Madelung constant, the remaining systematic
FS errors are expected to be dominated by periodic charge-image
quadrupole interactions, and to fall off rapidly as $L^{-3}$, where
$L$ is the linear size of the supercell.  Depending on whether
sufficient data are available, linear extrapolation in $1/L^3$ can be
used to remove these errors. One could even attempt to eliminate these
errors using the Makov-Payne expression for the correction to the
formation energy of a charged defect.\cite{makov1995}  For a 2D
material with a supercell size much less than $r_*$, the charge-image
quadrupole Keldysh interaction falls off as $1/L^2$; when the linear
size of the supercell exceeds $r_*$, a crossover to $1/L^3$ scaling
takes place.

The corrected $\qpg({\bf a}_1^{\rm sc},{\bf a}_2^{\rm sc},{\bf
  a}_3^{\rm sc})-v_{\rm M}({\bf a}_1^{\rm sc},{\bf a}_2^{\rm sc},{\bf
  a}_3^{\rm sc})$ data are also subject to additional,
beyond-linear-response effects.  These additional effects are
quasirandom, scaling in no systematic way with system size; however,
they do correlate with analogous charged-defect formation energies
evaluated at the DFT level: see Sec.\ \ref{subsub:diamond_si}.  We
interpret these errors as commensurability effects: oscillations in
the electron pair density arising from additional quasiparticles (in a
metallic system these would be Friedel oscillations) are artificially
made commensurate with the supercell.

Some earlier QMC studies have extrapolated gaps to infinite
system size assuming FS errors in energy gaps scale as $1/L$.\cite{ertekin2013,mostaani2016} For a fixed cell shape,
$v_{\rm M}({\bf a}_1^{\rm sc},{\bf a}_2^{\rm sc},{\bf a}_3^{\rm sc})$
itself scales like $1/L$, so this \textit{Ansatz} is reasonable.
However, this approach is invalid if the cell shape is varied.
Furthermore, it is difficult to extrapolate reliably from a small
number of data points suffering from unquantified quasirandom noise.
In many cases, averaging corrected energy gaps is a more accurate way
of removing systematic and quasirandom FS effects.  As shown
in Sec.\ \ref{subsub:diamond_si} (Table \ref{table:si_shapes_gaps},
specifically), the magnitude of the quasirandom FS effects
appears larger than any remnant systematic FS error after
application of our proposed correction [Eq.\ (\ref{eq:qpg_corr})] in
three-dimensional Si; in 2D phosphorene, however, residual systematic
FS errors are still present after the Madelung-constant
correction has been applied, as shown in
Sec.\ \ref{subsub:phosphorene}.  Whether extrapolation in $1/L^3$ or simple
averaging of corrected gaps is the most effective way of removing
FS effects depends on the system and on the number of system
sizes at which gap data are available.  In either case, provided the
quantified QMC statistical error bars are less than the unquantified
quasirandom FS noise (typically around 0.1 eV), the data
should not be weighted by the inverse square QMC error bars when
extrapolating or averaging.

For a fixed supercell size $N$, one can choose a cell shape to maximize the
distance
between periodic images, thereby minimizing remaining systematic FS
effects not accounted for by Eq.\ (\ref{eq:qpg_corr}).  For cubic materials,
the cells that maximize the nearest-image distance are themselves cubic
($n\times n\times n$ arrays of unit cells). In other lattice systems, the
supercells maximizing the nearest-image distance need not be of the same
shape as the primitive cell, or even be diagonal in their extent.
Nondiagonal supercells have previously been used in studies of lattice
dynamics at the DFT level,\cite{lloydwilliams2015} but purely as a
means of reducing computational expense.
The shape of the simulation supercell may also be of
significance with regards to the quasirandom FS effects: see
Sec.\ \ref{subsub:diamond_si}.

\begin{figure}[ht!]
  \centering
  \includegraphics[width=\linewidth]{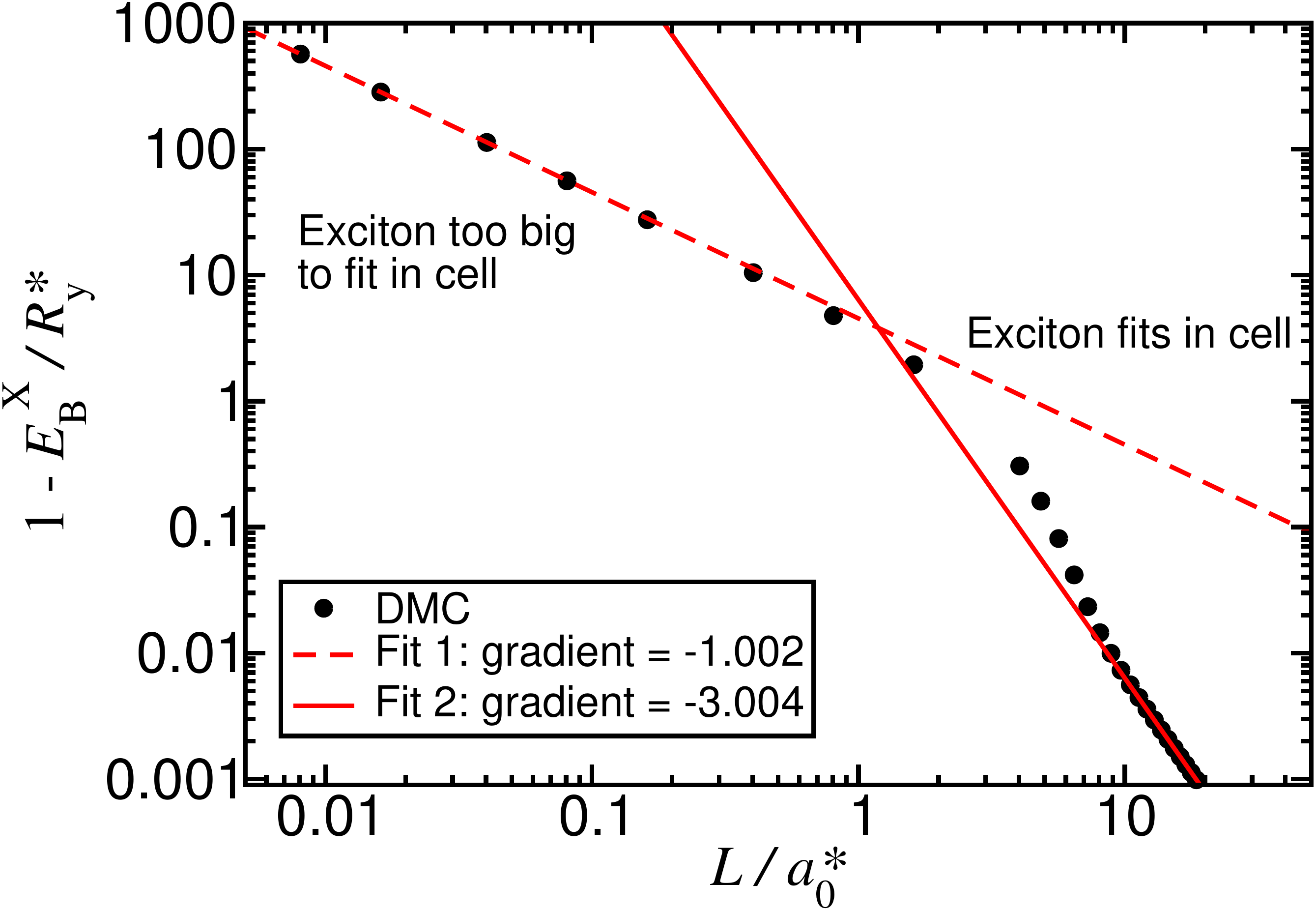}
  \caption{(Color online) Scaled difference of exciton binding energy
    $E_{\rm B}^{\rm X}$ and the exciton Rydberg against the lattice
    parameter $L$ in an effective-mass model of a three-dimensional
    exciton confined in a periodic FCC cell. $R^{*}_{\rm y}=m_{\rm
      e}^* m_{\rm h}^*/[2\epsilon^2(m_{\rm e}^*+m_{\rm h}^*)]$ and
    $a^{*}_{0}=\epsilon(m_{\rm e}^*+m_{\rm h}^*)/(m_{\rm e}^* m_{\rm
      h}^*)$ are the exciton Rydberg and the exciton Bohr radius,
    respectively, where $m_{\rm e}^*$ and $m_{\rm h}^*$ are the
    electron and hole masses and $\epsilon$ is the permittivity.
    The exciton Rydberg is the binding energy of an exciton in a
    cell of infinite extent.  The gradient on this log-log plot gives
    the scaling exponent of the finite-size error in the exciton
    binding energy.}
\label{fig:ex_v_3d}
\end{figure}

For the case of excitonic gaps, there is another FS effect to
consider.  The characteristic size of an exciton is usually the
exciton Bohr radius $a_{\rm B}^*=\epsilon/\mu$, where $\mu=m_{\rm
  e}^*m_{\rm h}^*/(m_{\rm e}^*+m_{\rm h}^*)$ is the electron-hole
reduced mass, $\epsilon$ is the permittivity, and $m_{\rm e}^*$ and
$m_{\rm h}^*$ are the electron and hole effective masses,
respectively. (Note that the size of an exciton is different in 2D
materials where the screened interaction is of Keldysh
form;\cite{Mostaani2017} in that case the size of the exciton is
$r_0=\sqrt{r_*/(2\mu)}$.) If the simulation supercells used are of
linear size much less than the characteristic exciton size then the
exciton is artificially confined and the kinetic energy dominates the
Coulomb interaction. The exciton consists of two weakly attracting,
almost independent quasiparticles, and the FS behavior of the
resulting ``excitonic'' gap mimics that of the quasiparticle gap, with
a FS error dominated by the Madelung energies of the free electron and
hole. If, on the other hand, the simulation supercell has a linear
size exceeding the characteristic size of the exciton, the
hydrogen-like bound state forms, and the leading-order systematic FS
scaling in the excitonic energy gap is given by the energy of a
lattice of self-image-interacting excitons. To investigate the binding
energy of a lattice of exciton images, we have performed a series of
two-particle DMC calculations in which an electron and a hole in the
effective mass approximation and interacting by the Ewald interaction
are confined to a face-centered cubic (FCC) cell of lattice parameter
$L$.  The results of this investigation are presented in
Fig.\ \ref{fig:ex_v_3d}, which clearly shows the crossover in the
scaling of the FS error in the exciton binding energy from $L^{-1}$ in
small cells to $L^{-3}$ in large cells when the linear size of the
cell is about twice the exciton Bohr radius. The 2D\cite{wood2004} and
3D\cite{makov1995,Fraser1996} Ewald interactions, $v_{\text{Ew}}(r)$,
may be expanded into the general form
\begin{equation}
  v_{\text{Ew}}(r) - v_{\rm M} = v_{\text{Coul}}(r) + a \dfrac{r^2}{L^3}
  + O(r^{4}), \label{eq:ewald_expansion}
\end{equation}
where $v_{\rm M}$ is the Madelung constant and $a$ is a geometrical
factor, which is sensitive to dimensionality and the supercell
shape. $v_{\text{Coul}}(r)$ is the aperiodic Coulomb interaction. This
difference from the exact Coulomb interaction is the physical source
of the $L^{-3}$ FS error in the exciton binding energy as evaluated in
calculations employing periodic boundary conditions.\footnote{In an
  anisotropic system, the quadratic term in
  Eq.\ (\ref{eq:ewald_expansion}) would be replaced by a bilinear form
  ${\bf r}^{\intercal} T {\bf r}$, with $T$ a tensor depending on the
  lattice structure.}  In a sufficiently large cell, the exciton wave
function is nearly independent of linear system size $L$, and hence by
first-order perturbation theory the effect of the $a r^2 / L^3$ term
goes as $L^{-3}$.  Once again, the situation is different in 2D
materials when the simulation supercell is much smaller than $r_*$
(but larger than the exciton size $r_0$); in that case the finite-size
error in the exciton binding energy and hence excitonic gap scales as
$L^{-2}$.

The approximate FS behavior of the excitonic gap is
determined by the FS behavior of the exciton binding energy
$E_{\rm B}^{\rm X}({\bf a}_1^{\rm sc},{\bf a}_2^{\rm sc},{\bf
  a}_3^{\rm sc})$.  In particular, the excitonic gap in a finite
supercell is approximately given by
\begin{equation}
\exg({\bf a}_1^{\rm sc},{\bf a}_2^{\rm sc},{\bf a}_3^{\rm sc}) \approx
\qpg(\infty)-E_{\rm B}^{\rm X}({\bf a}_1^{\rm sc},{\bf a}_2^{\rm sc},
{\bf a}_3^{\rm sc}).
\end{equation}
If the exciton Bohr radius is large compared with the supercell then
$E_{\rm B}^{\rm X}({\bf a}_1^{\rm sc},{\bf a}_2^{\rm sc},{\bf
a}_3^{\rm sc}) \approx -v_{\rm M}({\bf a}_1^{\rm sc},{\bf a}_2^{\rm
sc},{\bf a}_3^{\rm sc})$, so that the FS behavior of the
quasiparticle and excitonic gaps is the same, and either can be used
to estimate the infinite-system quasiparticle gap by subtracting the
screened Madelung constant from the result obtained in a finite
supercell.  There is no point in attempting to calculate exciton
binding energies using differences of quasiparticle and excitonic gaps
in supercells smaller than the exciton Bohr radius suggested by the
effective-mass approximation.  On the other hand, if the simulation
supercell is larger than the exciton Bohr radius then the FS
errors in the exciton binding and hence excitonic gap are small and
fall off rapidly as $L^{-3}$; in this case it is possible to determine
the exciton binding energy.

We have investigated whether single-particle FS effects
(i.e., momentum-quantization effects) are significant in DMC gaps by
fitting
$\Delta(N)=\Delta(\infty)+b/N^{1/3}+c[\Delta^{\text{DFT}}(N,{\bf
    k}_{\text{s}}) - \Delta^{\text{DFT}}(\infty)]$ to DMC gaps
$\Delta(N)$ obtained in a series of cells of the same shape but
different size $N$, where $\Delta^{\text{DFT}}(N,{\bf k}_{\text{s}})$
is the DFT energy gap evaluated for a finite supercell containing $N$
electrons, ${\bf k}_{\text{s}}$ is the offset to the grid of ${\bf
  k}$-vectors used in the DFT calculation, and
$\Delta^{\text{DFT}}(\infty)$ is the DFT gap converged with respect to
${\bf k}$-point sampling.  However, we do not find the fitted values
of $c$ to be statistically significant.  Nor do we find correlation
between the ground-state DFT total energy and the QMC gaps. On the
other hand, we do observe some correlation with FS effects in
DFT-calculated defect-formation energies (see Fig.\ \ref{fig:dft_vs_dmc_gaps}).
Twist averaging\cite{lin2001twist} (TA) is a
method for removing single-particle FS effects from ground-state
expectation values. TA involves averaging results over
simulation-supercell Bloch vectors ${\bf k}_{\rm s}$, i.e., over
offsets to the grid of ${\bf k}$ vectors. However, in gap calculations
the value of ${\bf k}_{\rm s}$ is fixed by the need to ensure that the
${\bf k}$ points involved in the excitation are present in the grid;
hence TA in the conventional sense cannot be used in QMC excitation
calculations.

\subsection{QMC band structures: dipole matrix elements and the spectral
function}\label{sub:qmc_bs}

Quasiparticle energies are generally complex quantities, because
quasiparticle excitations have finite lifetimes. The central quantity
of interest in many spectroscopic experiments is the spectral function
$A({\bf k},\omega)$, which characterizes the electronic states of wave vector
${\bf k}$ in a
given material, having peaks centered on the quasiparticle energies
$\omega$ whose widths relate to the lifetime of the quasiparticle
excitation in question. It would be possible to try to extract the
energy-momentum spectral function from VMC calculations. As an
example, one could calculate the squared matrix element
\begin{equation}
  \lvert \langle \Psi_{N}({\bf r}_1,\ldots,{\bf
    r}_N)\cdot\exp{\left[i{\bf k}\cdot {\bf r}_{N+1}\right]} |
  \Psi_{N+1}({\bf r}_{1},\ldots,{\bf r}_{N+1}) \rangle \rvert^2,
\end{equation}
for the HEG at the VMC level, where $\Psi_{N}$ is an optimized
$N$-electron wave function. This would allow for determination of the
broadening of the spectral peak at a particular momentum ${\bf k}$ and
extraction of the lifetime of quasiparticles in the quasielectron band
at ${\bf k}$, complementing previous works. This would go some way to
completing the first-principles description of the properties of the
HEG from the point of view of Landau's Fermi liquid
theory.\cite{landau1957theory,landau1957oscillations,landau1959theory}

A similar possibility would be to try to calculate the radiative
lifetime for an excitonic state. This relies on the evaluation of
dipole matrix elements, which again is possible with VMC\@. This has
already been performed for few-body systems in a simple
model,\cite{heterobilayer} and for the 2$^2$S $\rightarrow$ 2$^2$P transition
of the Li atom.\cite{barnett1992monte}

One might think that a natural way to obtain improved estimates of
quasiparticle lifetimes and radiative rates would be to evaluate the
corresponding matrix elements at the DMC level. However, this is not
immediately possible. The DMC method gives no direct information
regarding many-electron wave functions [i.e., produces no functional
form for $\Psi_N({\bf R})$].\footnote{The distribution which the DMC
algorithm samples is either the ``mixed'' distribution (the product
of the fixed-node ground state with the trial wave function) or, if
the future-walking algorithm is used,\cite{barnett1991monte} the ``pure''
distribution (the modulus square of the fixed-node ground state
itself). This does not change the salient point, which is that DMC
{\it generates configurations\/} and does not supply a functional
form for the fixed-node ground state.}

\subsection{Excitations in metallic systems}\label{sub:exc_metals}

Various studies have investigated, from a microscopic viewpoint, the
excited-state properties of the 2D
HEG\@.\cite{kwon1994,holzmann2009,drummond2013diff,drummond2013}
This involves the study of intraband excitations, in which electrons are
promoted or added into higher energy states on the free-electron-like band of
the HEG in order to determine the quasiparticle effective mass and the Fermi
liquid parameters. All of these studies have observed the presence of severe
finite-size effects. In what remains of the present article, we will discuss
only interband excitations to calculate energy gaps.

\subsection{Computational expense}

Methods developed to improve the scaling of QMC
calculations\cite{alfe2004linear,williamson2001} may find use in
excitation calculations. By localizing low-lying states which are not
directly involved in excitations, the number of nonzero orbitals to
evaluate at a given point ${\bf r}$ is reduced, and the Slater matrix
is made sparse, improving the cost scaling of the Slater part of the
wave function by a factor of $N$. An additional side effect of this is
to reduce the computational expense of the inclusion of backflow
correlations (whose dominant cost arises at the orbital-evaluation
stage of a calculation). However, a major problem with the use of
localized orbitals is that, in order to obtain efficiency increases,
one sacrifices accuracy in individual total energies by truncating
localized orbitals to zero at finite range. The extent to which this
loss of accuracy will affect total-energy differences in solids is
unclear, although early studies on molecules have provided positive
results.\cite{williamson2002} Given that other biases (single-particle
finite-size effects, time-step bias, etc.)\ cancel so well in gap
calculations in solids (see Sec.\ \ref{subsub:diamond_si}) we expect
the loss in accuracy in energy gaps due to the truncation of localized
low-lying electronic states to be very small.  On the other hand,
computational expense is often dominated by other factors such as the
evaluations of two-body terms in the Jastrow factor and updates to
the Slater matrix, limiting the scope for speedup.

Because highly precise total energies are required from the DMC
calculations used in forming energy gaps, the most significant portion
of computational time is spent in the statistics-accumulation phase;
the equilibration phase is only a small fraction of the total
computational expense. This means that QMC gap calculations are
particularly suited to massively parallel computational architectures.

\subsection{Nuclear relaxation and vibrational
effects}\label{sub:vibrational_effects}

The renormalization of
static-nucleus energy gaps by zero-point vibrational effects is
important for any comparison of theoretical results with
experiment.\cite{monserrat2014} In the extreme case of hexagonal ice,
this effect contributes a correction in the range of 1.5--1.7
eV\@.\cite{engel2015vibrational,engel2016vibrational} Related work has
also demonstrated a large renormalization of the energy gap in the
benzene molecule by more than 0.5 eV\@.\cite{mostaani2016}
We investigate this issue in Sec.\ \ref{subsub:h2_dimer},
where we present results for an H$_2$ molecule with a full quantum
treatment of both protons and electrons.

A second issue is the equilibrium geometry of electronic excited states.  In an
adiabatic ionization potential, electron affinity, or quasiparticle gap, the
geometry of the molecule or crystal is allowed to relax after the addition or
removal of an electron.  By contrast, in a ``vertical'' ionization potential,
electron affinity, or quasiparticle gap, the atomic structure of the cation or
anion is assumed to be the same as that of the ground state.  An important
point to note here is that, from the point of view of experiment, atomic
relaxation may or may not be relevant.  Experimental measurements that occur on
timescales smaller than those associated with the structural relaxation of a
molecule or a solid (for example, as with photoemission/inverse photoemission
spectroscopy) are insensitive to any relaxation effects which are instigated by
the measurement.  On the other hand, in experimental measurements that occur on
timescales greater than those associated with the structural relaxation (for
example, as in zero electron kinetic energy spectroscopy\cite{Muller1991}), one
can expect that one will measure directly an adiabatic excitation energy, and
that comparison to fully relaxed \textit{ab initio} results is reasonable. The
situation is less clear in the case that the experimental and structural
relaxation timescales are comparable.  Geometrical relaxation in excited states
typically reduces quasiparticle gaps by 0.1--0.5 eV\@. We present many of our
quasiparticle-gap results with and without relaxation in excited states, using
DFT to relax structures.  A closely related issue is the Stokes shift, which is
the difference between excitonic absorption and emission gaps.  In an
absorption gap, the geometry is that of the ground state; in an emission gap,
the geometry is that of the excited state.  QMC calculations have previously
been performed to calculate Stokes shifts in diamondoids using DFT
geometries.\cite{Marsusi2011}

Both of these issues complicate the detailed comparison of \textit{ab
  initio} gaps with experimental measurements.

\section{Computational details\label{sec:methodology}}

\subsection{DFT orbital generation}\label{sub:orbital_generation}

Our DFT calculations were carried out with the \textsc{castep}
plane-wave-basis code.\cite{Clark2005} In the case of molecules and of
phosphorene, prior to any wave-function generation calculation, we
relaxed the ground-state (and, where explicitly stated, excited-state)
geometries to within a force tolerance of at most 0.05 eV/\AA, with
ultrasoft pseudopotentials\cite{vanderbilt1990soft} representing the
nuclei and core electronic states. All of our DFT calculations used
the Perdew-Burke-Ernzerhof (PBE) parameterization of the generalized
gradient approximation to the exchange-correlation
energy.\cite{Perdew1996} For our calculations on solids, we used
experimentally obtained geometries (Si from
Ref.\ \onlinecite{CODATA2014}, cubic boron nitride (BN) from
Ref.\ \onlinecite{Goncharov2007}, and $\alpha$-SiO$_2$ from
Ref.\ \onlinecite{Levien1980}).

We have used Trail-Needs Dirac-Fock averaged-relativistic-effect
pseudopotentials\cite{Trail2005,Trail2005a} for all wave-function
generation calculations and subsequent QMC calculations, except in
our all-electron calculations. We have chosen the local channels of
our pseudopotentials such that no ghost states exist, and we have used
plane-wave cutoff energies which lead to an estimated DFT basis-set
error per atom of at most 10$^{-4}$ a.u.\ (2.72 meV)\@.\cite{Drummond2016}

After their generation, the DFT single-particle orbitals were
rerepresented in a blip (B-spline) basis.\cite{Alfe2004} This
allows for improved computational efficiency of QMC calculations, and
the removal of unphysical periodicity in calculations on zero-, one-,
and two-dimensional systems.

\subsection{QMC calculations}\label{sub:qmc_calculations}

\subsubsection{Slater-Jastrow(-backflow) wave
functions}\label{subsub:slater_jastrow_wfns}

We have used Jastrow factors of the form outlined in
Ref.\ \onlinecite{drummond2004jastrow} in all of our QMC calculations,
with system-appropriate terms and with free parameters optimized by
unreweighted variance minimization and subsequent energy
minimization.\cite{umrigar1988opt,drummond2005var,toulouse2007opt,Umrigar2007}
We have not (except where explicitly stated) reoptimized
Jastrow-factor parameters in trial excited states.  We have used
backflow functions of the form outlined in
Ref.\ \onlinecite{lopezrios2006inhomogeneous}, optimizing free
parameters by energy minimization.\cite{Umrigar2007}

The results of our DMC calculations have been simultaneously
extrapolated to infinite population size, and zero time step in an
efficient manner.\cite{Lee2011} We have used the ``T-move'' method of
Casula to ensure that our DMC energies are variational in the presence
of nonlocal pseudopotentials.\cite{Casula2006} All of our QMC
calculations have been carried out using the \textsc{casino}
code.\cite{needs2009continuum}

\subsubsection{Multideterminant trial wave functions}\label{subsub:mdets}

In a multideterminant wave function, the Slater part of the wave
function of Eq.\ (\ref{eq:sjwfn}) is replaced by
\begin{equation}
\mathcal{D}({\bf R})
\rightarrow \mathcal{D}({\bf R}) + \sum_{j}c_j\mathcal{D}_j({\bf R}),
\end{equation}
where the original determinant $\mathcal{D}$ is chosen as the
``dominant'' determinant, and the excited determinants $\mathcal{D}_j$
are populated with single-particle orbitals with substituted
degenerate or near-degenerate orbitals of interest with respect to
those appearing in $\mathcal{D}$. Unless one believes the
single-particle theory used to generate the orbitals to be
qualitatively incorrect, the order of the eigenvalues of the orbitals
occupied in the Slater determinant of single-particle orbitals is
preserved with respect to the interacting case: the states of the
interacting and noninteracting systems are assumed to be adiabatically
connected. In the case of a failure of the single-particle theory,
this is not guaranteed, and the state formed from the determinant of
single-particle orbitals is not a reasonable trial state.  E.g., in a
case where DFT metallizes an insulator, one might attempt to remedy
the problem by, e.g., inclusion of exact exchange (the use of a hybrid
functional, or even Hartree-Fock theory itself) or artificial
separation of the occupied and unoccupied manifolds (i.e., the use of
a scissor correction) in the orbital-generation calculation.

One is able to obtain better estimates of ground-state total energies
by variation of the multideterminant expansion coefficients $\{c_j\}$.
One might also be able to obtain better estimates of certain
excited-state energies (see Sec.\ \ref{sub:qp_ex_gaps}). However,
general excited states do not obey variational principles, and so it
is not obviously the case that one would always want to form a
multideterminant expansion for the excited state.

There are cases where the formation of a (restricted) multideterminant
expansion is desirable. Firstly, excited-state multideterminant expansions
transforming as 1D irreps of the full
symmetry group of the Hamiltonian of a system can be shown to obey variational
principles in fixed-node DMC,\cite{Foulkes1999} as discussed in Sec.\
\ref{sub:wfn_nodes_vps}. Secondly, in cases of states with degeneracy or
near-degeneracy, one might expect that the wave function should have some
multireference character. Such degeneracies are much more likely to occur in
the excited state than in the ground state. The inclusion of determinants
characterizing electron promotions (or additions, or removals) from the
degenerate or near-degenerate energy levels might reduce excited-state
energies, leading to lower QMC energy gaps. Towler \textit{et
al.}\cite{Towler2000} paid a great deal of attention to the correct inclusion
of degenerate determinants of specified symmetry classes in their study of
diamond (which has the same symmetry properties as Si,
with the same consequence that the valence-band maximum and conduction band
at $\Gamma$ are triply degenerate at the single-particle
level). When choosing a multideterminant expansion to describe an excited
state, one must apply a group theoretical projection operator to each of the
possible degenerate determinants in order to determine an excited-state trial
wave function of definite symmetry. This ``safe'' trial wave function is then a
few-determinant expansion in the space of degenerate determinants of
single-particle orbitals, with a definite symmetry. However, this symmetry may
only be maintained at the VMC level, and the fixed-node DMC algorithm may still
break it if the trial wave function does not transform as a 1D
irrep.  The weaker variational principle for DMC excited
states mentioned in Sec.\ \ref{sub:wfn_nodes_vps} still applies in
cases where trial functions have specific transformation
properties, however.

We have explicitly tested the formation of multideterminant trial wave
functions in some of our calculations in Si (see
Sec.\ \ref{subsub:diamond_si}), where three bands at the $\Gamma$
point are degenerate in the absence of spin-orbit coupling.

\section{Results and discussion\label{sec:results}}

\subsection{Atoms\label{sub:atoms}}

\subsubsection{H atom: a model of excited-state fixed-node
errors}\label{subsub:h_nodes}

An important class of fixed-node errors in excited-state DMC
calculations is that which may arise due to the lack of a variational
principle. Here we consider various modifications to the hydrogenic 2s
orbital, whose exact energy is $-\frac{1}{8}$ a.u. The corresponding
wave function is isotropic and hence transforms as the trivial 1D
irrep of the SO(3) geometric symmetry group of
the H atom; however, it is not the lowest energy eigenfunction of this
symmetry.  The nodal surface of the 2s orbital is a sphere of radius 2
a.u.  This example was previously investigated analytically in
Ref.\ \onlinecite{Foulkes1999}; here we provide numerical results that
corroborate the argument in Ref.\ \onlinecite{Foulkes1999}, and we
investigate the consequences for optimization of backflow functions in
excited states.

\begin{figure}[ht!]
  \centering
  \includegraphics[width=\linewidth]{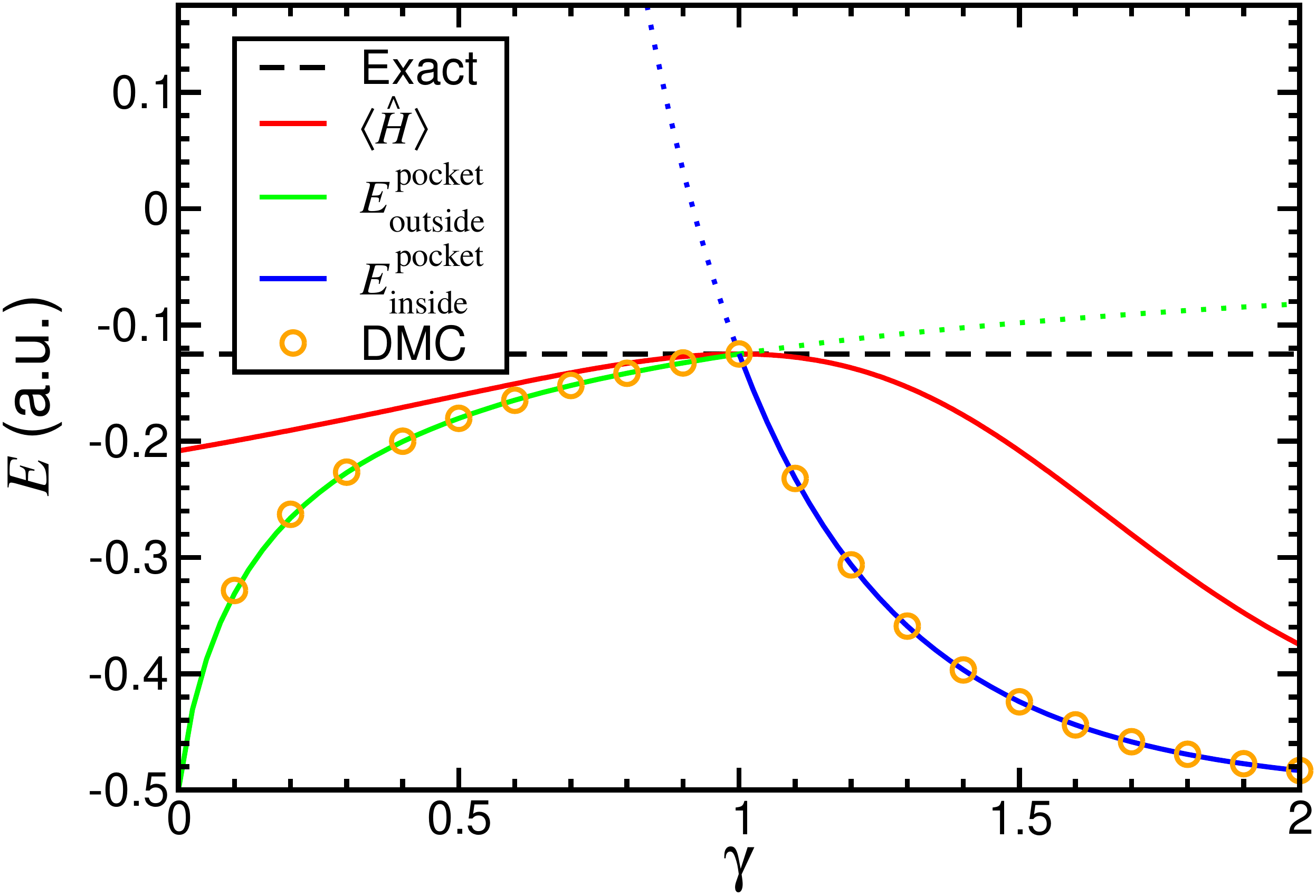}
  \caption{(Color online) Approximations to the first-excited-state
    energy of an H atom using the $\psi^{\gamma}$ excited-state trial
    wave function of Eq.\ (\ref{eq:node_vol}) as a function of
    $\gamma$, obtained by various means. DMC errors are smaller than
    the thickness of the lines. The pocket eigenvalues outside and
    inside the nodal surface, $E_{\rm outside}^{\rm pocket}$ and
    $E_{\rm inside}^{\rm pocket}$, were determined by numerical
    integration of the Schr\"{o}dinger equation with fixed-node
    boundary conditions, and $\langle \hat H \rangle =
    \langle\psi^{\gamma} \lvert \hat H \rvert \psi^{\gamma} \rangle$,
    where $\hat H$ is the Hamiltonian.}\label{fig:h_radial_nodes}
\end{figure}

\begin{figure}[ht!]
  \centering
  \includegraphics[width=\linewidth]{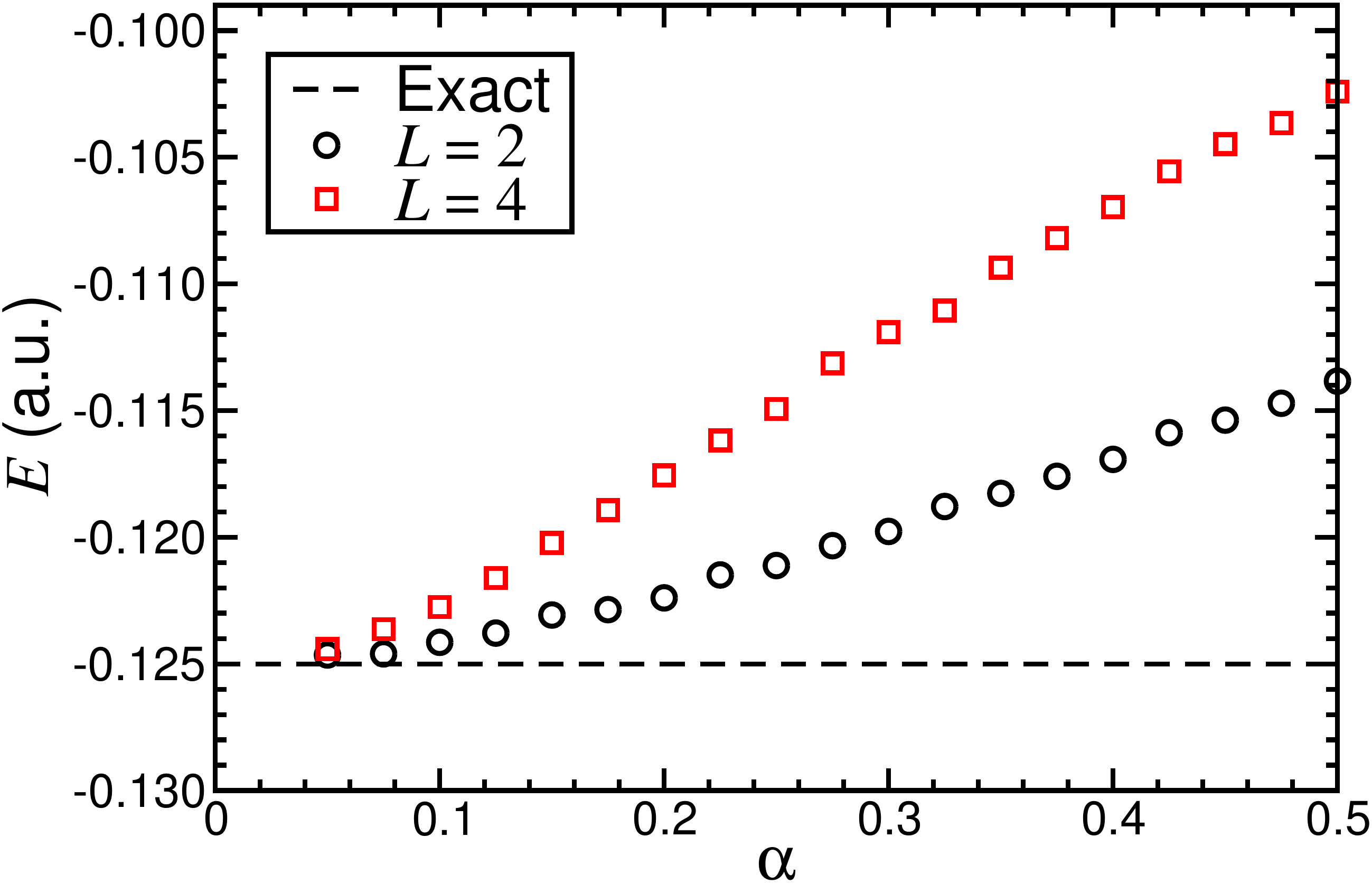}
  \caption{(Color online) DMC first-excited-state energies of an H
    atom with the trial wave function $\psi^{\alpha,\beta}_L$ [see
      Eq.\ (\ref{eq:wrinkle_node})] for $L=2$ and $4$ at various
    amplitudes $\alpha$ of wrinkling of the nodal surface [see
      Eq.\ (\ref{eq:wrinkle_node})]. DMC error bars are of order the
    size of the symbols.}\label{fig:h_alpha_plot}
\end{figure}

The two ways that a spheroid nodal surface can be inexact are that (a)
the average positions of nodes is incorrect, and/or (b) the curvature
of nodes is incorrect. We have studied two inexact nodal surfaces for
the 2s state using the trial wave functions
\begin{align}
  \psi^{\gamma}(r) =&\ C_{\gamma}\left( 2\gamma - r\right)
  \exp{\left(-\frac{r}{2}\right)}, \label{eq:node_vol} \\
  \psi^{\alpha,\beta}_L(r,\theta) =&\ D_{\alpha,\beta}
  \left\{ 2\beta\left[1+\alpha\mathcal{Y}_{L,0}(\theta) \right] -
  r\right\} \exp{\left(-\frac{r}{2}\right)},
\label{eq:wrinkle_node}
\end{align}
which are exact (2s) eigenstates for $\gamma = 1$ and $\alpha=0$,
$\beta=1$.  The wave function $\psi^{\gamma}(r)$ encodes the scenario
already explored in Ref.\ \onlinecite{Foulkes1999}. The normalization
constants $C_{\gamma}$ and $D_{\alpha,\beta}$ are irrelevant in DMC,
and $\mathcal{Y}_{L,m_L}$ is a spherical harmonic. We have used
$\psi^{\gamma}$ as a DMC trial wave function with $\gamma$ being a
control parameter which varies the nodal volume, keeping the node
spherical. This addresses point (a). We have also used
$\psi^{\alpha,\beta}_L$ as a DMC trial wave function, with $\alpha$ a
control parameter that sets the degree of nonspherical distortion of
the nodal surface, this time with $\beta$ chosen to fix the nodal
volume to the exact value. This addresses point (b). The nodal
topology of our trial wave function does not change as a function of
$\gamma$ and $\alpha$; there are always two nodal pockets. The results
of varying $\gamma$ and $\alpha$ are presented in
Figs.\ \ref{fig:h_radial_nodes} and \ref{fig:h_alpha_plot}.

Define the pocket eigenvalues $E_{\rm outside}^{\rm pocket}$ and $E_{\rm
inside}^{\rm pocket}$ to be the energy eigenvalues associated with single
electrons occupying the regions outside and inside the nodal surface of
$\psi^\gamma$, respectively, where the boundary conditions are that the pocket
eigenfunctions are zero outside of their respective pockets. For the first
case, the pocket eigenvalues can be determined via numerical solution of a
model eigenvalue problem.  If the radial Schr\"{o}dinger equation is
integrated, but with a ``nodal boundary condition'' $\psi^{\gamma}(2\gamma)=0$,
then the lower of the corresponding eigenvalues $\min\{E_{\rm otside}^{\rm
pocket},E_{\rm inside}^{\rm pocket}\}$ matches very closely the DMC energy.
Moreover, we can also find the pocket eigenvalues corresponding to solutions
inside and outside the nodal surface for all $\gamma$ (see extended dotted
lines; only the lesser of these solutions is sampled by the DMC algorithm).
Even in the $\gamma \rightarrow 0$ and $\gamma \rightarrow \infty$ nodeless
limits the ground-state variational principle is always obeyed, i.e.,
$E\ge-\frac{1}{2}$ a.u.

There is a qualitative difference in the behavior of the energy expectation
value  $\langle \hat{H} \rangle$ (which could be evaluated by VMC)
versus the fixed-node DMC energy as a function of $\gamma$: the error in the
DMC excited-state energy due to the use of an inexact nodal surface is more
severe, and is first-order in the error in the nodal surface (as quantified by
$\gamma$).  Recall that the fixed-node error in the DMC ground-state energy
is second order in the error in the trial nodal surface.

In the second case, as is shown in Fig.\ \ref{fig:h_alpha_plot}, the fixed-node
error is always positive for $\alpha \neq 0$.  This is not too surprising,
given that if the wave function is to satisfy the nodal constraint, it must
adopt additional curvature in both nodal pockets. Additional curvature in space
corresponds to an increased kinetic energy of the wave function in both nodal
pockets.  The fixed-node DMC energy is second-order in the parameter $\alpha$,
because it is an even function of $\alpha$.

This model serves as an illustrative example of the fact that
excited-state fixed-node errors can be either positive or negative,
depending on the nature of the inexactness of the nodal surface. This
is important, in particular, if one is to attempt to improve the nodal
surface in a trial excited state. Even if the optimizable parameters
of a trial excited-state wave function cannot change the nodal
topology, optimization by energy minimization may result in the
development of a pathological nodal surface that gives a DMC energy
that is too low.

\begin{figure}[ht!]
  \centering
  \includegraphics[width=0.7\linewidth]{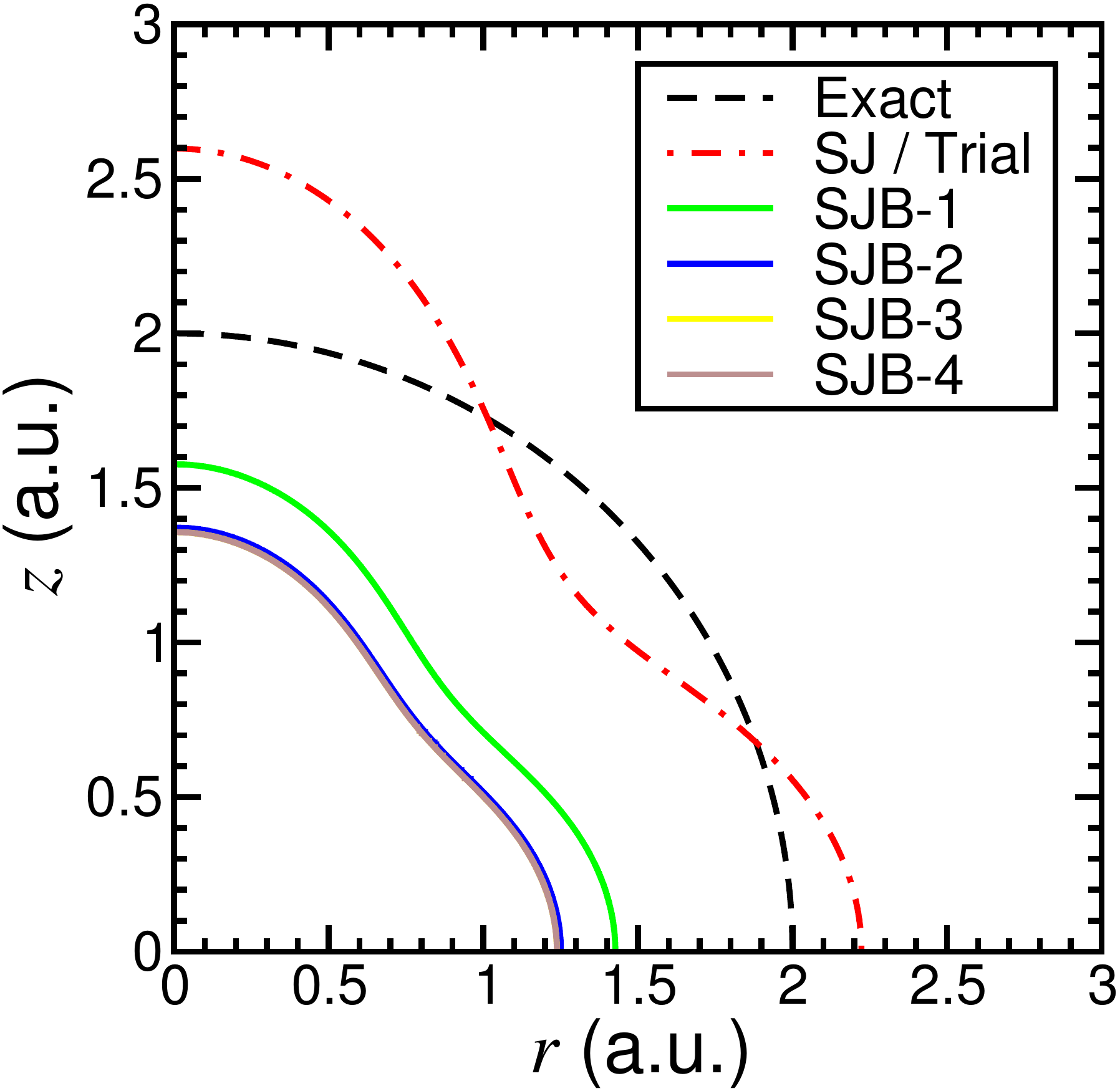}
  \caption{(Color online) Nodal surface of the SJB trial wave function
    $\psi^{\alpha=0.1,\beta}_{L=4}$ [see Eq.\ (\ref{eq:wrinkle_node})]
    for the first excited state of an H atom. The wave function is
    optimized by VMC energy minimization. SJB-$n$ labels the nodal
    surface of the SJB wave function after the $n^{\text{th}}$ cycle
    of energy minimization. The $n=3$ and $4$ cases are
    indistinguishable from each other, and correspond to the
    termination of the optimization process.}\label{fig:h_nodes_plot}
\end{figure}

We have tested this explicitly for the case of a trial wave function
$\psi^{\alpha=0.1,\beta}_{L=4}(r,\theta)$, with an electron-nucleus backflow
function. Successive cycles of energy minimization lower the VMC energy of this
state from $-0.1180(2)$ a.u.\ ($> -\frac{1}{8}$ a.u., positive error) to
$-0.1445(3)$ a.u.\ ($< -\frac{1}{8}$ a.u., negative error). This is exacerbated
at the DMC level, where the energy of the state with the optimal backflow
function drops further still to $-0.1562(3)$ a.u.  Throughout VMC optimization,
the nodal surface alters significantly, as shown in Fig.\
\ref{fig:h_nodes_plot}.

This investigation of the hydrogen atom suggests that the lack of
variational principle for excited-state energies is only a significant
problem if one attempts to reoptimize a parameter that moves the nodal
surface in an excited state.

\subsubsection{Ne atom: VMC and backflow?\label{subsub:neon}}

In terms of computational cost, VMC is several times cheaper than
DMC\@. It would therefore be desirable to know whether or not energy
gaps at the VMC level can be of comparable quality to their DMC
counterparts. To this end we have calculated the $n^{\text{th}}$
ionization potential of all-electron Ne up to and including $n=8$, at
various levels of theory (SJ-VMC, SJB-VMC, SJ-DMC, and SJB-DMC)\@. It
has previously been shown that SJB-VMC is capable of retrieving large
fractions (more than 99\%) of the correlation energy (defined with
respect to the then-best SJB-DMC energy) of the Ne and Ne$^+$
species;\cite{drummond2006quantum} however, no attempt was made to
evaluate the effectiveness of this approach beyond $n=1$. Our results
for the Ne atom are given in Table \ref{table:neon_ips}, alongside
corrected nonrelativistic literature
values.\cite{chakravorty1993ground}

\begin{table*}[!]
\centering
\caption{$n^{\text{th}}$ ionization potential of an all-electron Ne
  atom at various levels of QMC theory, together with corrected
  nonrelativistic experimental values.\cite{chakravorty1993ground} The
  mean absolute errors (MAEs) have been calculated over all ionization
  potentials obtained within a given level of theory.}\label{table:neon_ips}
\begin{ruledtabular}
\begin{tabular}{l *{5}{d{3.5}}|*{4}{d{3.5}}}
\multirow{2}{*}{$n$} & \multicolumn{5}{c|}{Ionization potential (eV)} &
\multicolumn{4}{c}{Error in ionization potential (eV)} \\
& \mc{Exact} & \mc{SJ-VMC} & \mc{SJB-VMC} & \mc{SJ-DMC} &
\multicolumn{1}{c|}{SJB-DMC} & \mc{SJ-VMC} & \mc{SJB-VMC} & \mc{SJ-DMC} &
\mc{SJB-DMC} \\ \hline
1& 21.61333&22.08(2)&21.96(2) &21.72(1) &21.72(1) &0.465&0.350&0.104&0.109\\
2& 40.99110&41.48(2)&41.39(2) &41.10(1) &41.06(1) &0.590&0.397&0.108&0.074\\
3& 63.39913&63.44(2)&63.23(1) &63.35(2) &63.39(1) &0.037&-0.173&-0.050&-0.010\\
4& 97.29312&97.91(2)&97.78(1) &97.75(2) &97.72(1) &0.616&0.489&0.458&0.424\\
5&126.28846&126.85(2)&126.72(2)&126.85(1)&126.79(1)&0.565&0.436&0.564&0.504\\
6&157.80001&158.43(2)&158.30(1)&158.25(2)&158.34(1)&0.630&0.496&0.453&0.545\\
7&207.04137&204.48(2)&204.56(1)&205.04(2)&205.26(1)&-2.561&-2.477&-2.005&
-1.786\\
8&238.78949&238.10(1)&238.49(1)&238.70(2)&238.79(1)&-0.687&-0.303&-0.089&
0.002\\
 \hline
 \mc{\text{MAE}} & \mc{0\%} & 0.83\%    & 0.67\%    & 0.38\%    & 0.34\%
\end{tabular}
\end{ruledtabular}
\end{table*}

As can be seen, the DMC ionization potentials match very closely the ``exact''
nonrelativistic results. The general trend that more sophisticated levels of
theory capture more of the correlation energy in excited states is observed, in
that the MAE follows the expected trend: SJ-VMC does very well, SJB-VMC
does better, SJ-DMC does better still, and SJB-DMC is our best method.  In this
case, the system is absent of vibrational effects and relativistic effects have
been removed from the experimental data.  Hence the major source of error in
the DMC calculations is fixed-node effects. To test the impact of fixed-node
error on our ionization potentials, we have performed a test calculation with a
SJB wave function which was reoptimized in the Ne$^{+}$ cationic state.
Ionization potentials are differences in ground-state energies for different
numbers $N$ of electrons, and hence fixed-node error is always positive in each
of the two energies involved in forming the difference.  We find that the
SJB-VMC and SJB-DMC first ionization potentials are $21.51(1)$ eV and
$21.73(1)$ eV, respectively. The SJB-DMC first ionization potentials with and
without reoptimization are consistent with each other.  On the other hand, the
SJB-VMC first ionization potentials with and without reoptimization are
$21.51(1)$ eV and $21.96(2)$ eV respectively [with MAE values of $0.47(6)$\%
and $1.62(8)$\%], and here we see the most improvement from reoptimization. The
MAE of the SJB-DMC result is $0.52(5)$\%, meaning that the results from SJB-VMC
and SJB-DMC with reoptimized backflow functions are effectively as good as each
other---although SJB-VMC underestimates and SJB-DMC overestimates the
ionization potential.

A recent coupled cluster [CCSD(T)] calculation determined the first
and second ionization potentials of Ne as 21.564 eV and 44.3 eV,
respectively [absolute errors of 0.04930 eV (0.23\%) and 3.30890 eV
(8.1\%) with respect to the ``exact'' nonrelativistic results that we
have compared against].\cite{White2015} A less recent configuration
interaction calculation determined the eighth ionization potential of
Ne as 238.78440 eV [absolute error of 0.00509 eV
(0.0021\%)].\cite{Chung1991}

\subsection{Molecules\label{sub:molecules}}

\subsubsection{H$_2$ dimer\label{subsub:h2_dimer}}

We have evaluated the SJ-DMC first ionization potential of the H$_2$ dimer
using orbitals expanded in plane-wave and Gaussian
basis sets.  Our plane-wave calculations employed Trail-Needs
pseudopotentials, while our Gaussian basis set calculations were
all-electron. In our all-electron calculations, we have used bond
lengths matching the G2 values.\cite{curtiss1991gaussian} In the
pseudopotential calculations, we have relaxed geometries in the ground
(and excited, where specifically mentioned) states in DFT with the use
of the PBE exchange-correlation functional.

We have also carried out plane-wave-basis all-electron
calculations, where the full Coulomb interaction was used to evaluate the DFT
total energy. Such calculations are prohibitively expensive for atoms beyond C,
requiring very large plane-wave cutoff energies to achieve reasonable
convergence of total energies. We have carried out total-energy convergence
tests for this system, the results of which informed our choice of plane-wave
cutoff in orbital-generation calculations (500 a.u.). We estimate the error in
DFT total energies due to this choice of plane-wave cutoff energy to be $\sim
2\times10^{-3}$ a.u., and much smaller in DMC (where cusp
corrections\cite{kato1957eigenfunctions}
act to correct the wave function behavior at short range, which is the most
difficult region to represent in a plane-wave basis).
Our findings are displayed alongside experimental and other theoretical
estimates in Table \ref{table:o2_h2_gaps}.

\begin{table}[ht!]
\centering
\caption{DMC ionization potentials of the H$_2$ and O$_2$ dimers.
  All-electron (AE) and pseudopotential (PP) calculations have been
  performed with Gaussian (G) and plane-wave (PW) bases.  Calculations
  employing relaxed excited-state geometries are denoted ``ER\@.''
  The ``J-DMC (p$^+$p$^+$e$^-$e$^-$)'' calculations used a Jastrow wave
  function to describe
  the ground state of two distinguishable quantum protons and two
  distinguishable electrons for parahydrogen H$_2$, and the ground
  state of two distinguishable protons and one electron for the
  parahydrogen cation H$_2^+$.  Self-consistent quasiparticle $GW$
  results are denoted ``QS$GW$,'' coupled cluster results with single, double,
  and (triple) excitations ``CCSD(T)'' (``EPT'' means electron propagator
  theory), second-order M{\o}ller-Plesset
  perturbation theory results ``MP2,'' quadratic configuration interaction
  ``QCI'' (with levels of excitations as with coupled cluster), and results
  obtained by means of the generalized James-Coolidge expansion
  ``JCE\@.''}\label{table:o2_h2_gaps}
\begin{ruledtabular}
\begin{tabular}{lcc}
\multirow{2}{*}{Method} & \multicolumn{2}{c}{Ionization potential (eV)}\\
   & H$_2$ & O$_2$ \\
\hline
SJ-DMC (AE-PW)    & 16.465(3) & --  \\
SJ-DMC (AE-G)    & 16.462(6) & 13.12(7) \\
SJ-DMC (PP-PW)    & 16.377(1) & 12.84(2) \\
SJ-DMC (PP-PW-ER) & 15.582(1) & 12.33(2) \\
\hline
J-DMC (p$^+$p$^+$e$^-$e$^-$) & 15.4253(7)& -- \\
\hline
QS$GW$ & 16.04,\cite{Kaplan2016} 16.45\cite{Koval2014} & -- \\
CC-EPT & -- & 12.34,12.43\cite{Ortiz1992} \\
MP2 & -- & 11.72\cite{Su2011} \\
CCSD & -- & 11.76,12.13\cite{Stanton1992} \\
CCSD(T) & -- & 11.95\cite{Stanton1992} \\
QCISD(T) & -- & 12.18\cite{Su2011} \\
JCE & 15.42580\cite{Kolos1994} & -- \\
\hline
Experiment & 15.4258068(5)\cite{McCormack1989} & 12.0697(2)\cite{Tonkyn1989}
\end{tabular}
\end{ruledtabular}
\end{table}

It is clear that the use of pseudopotentials has some bearing on the
quality of the excitation results, but also that structural and
vibrational effects are critically important, as evidenced by the
strong reduction of the ionization potentials upon relaxation of the
excited-state geometry.

Experimental zero-point energies suggest that a reduction in the
calculated ionization potential of H$_2$ of around 0.02 eV is appropriate
to properly allow for
comparison with experiment.\cite{irikura2007} This is not enough to fully
bridge the gap between our best SJ-DMC results and the experimental ones.
However, we have used DFT-derived geometries, and have already shown that the
use of pseudopotentials incurs an error of order the remaining difference
between the (pseudopotential) SJ-DMC and experimental ionization potential.

For the simple case of a parahydrogen H$_2$ molecule (i.e., a molecule
with opposite-spin protons) it is feasible to perform DMC calculations
in which both the protons and electrons are treated as distinguishable
quantum particles.  Since the ground states of both the parahydrogen
molecule H$_2$ and the parahydrogen cation H$_2^+$ are nodeless, the
fixed-node DMC calculations are exact nonrelativistic calculations (in
the limit of zero time step, etc.).  We find the J-DMC total energies
of parahydrogen H$_2$ and the parahydrogen cation H$_2^+$ to be
$-1.16401(2)$ and $-0.5971396(3)$ a.u., respectively.\footnote{To
  extrapolate the J-DMC H$_2$ energy $E(\tau)$ to zero time step we
  used nine different time steps $\tau$, ranging from 0.0005 a.u.\ to
  0.032 a.u., and we found the time-step bias to consist of a
  crossover between two different linear regimes.  This is because
  there are two small length scales in the problem: the Bohr radius
  and the root-mean-square displacement of the protons in their
  vibrational ground state. We therefore performed the time-step
  extrapolation by fitting the Pad\'{e} form
  $E(\tau)=[E(0)+a\tau+b\tau]/(1+C\tau)$ to our data, where $E(0)$,
  $a$, $b$, and $C$ are fitting parameters.  We recommend this form of
  time-step extrapolation in other DMC calculations in which there is
  a separation of length scales that results in a crossover between
  two linear-bias regimes.}  As shown in Table \ref{table:o2_h2_gaps},
the resulting ionization potential then agrees with experiment to
within 0.01 eV\@. Another experimental study was able to resolve a
para-ortho splitting of 19(9) $\mu$eV in the ionization potential, and
determined the first ionization potential of parahydrogen specifically
as 15.425808(6) eV,\cite{Glab1987} a value which is consistent with
the averaged result of Ref.\ \onlinecite{McCormack1989}.

The results shown here demonstrate the critical importance of nuclear
geometry and vibrational effects on energy gaps on a subelectronvolt
scale.  To obtain excellent agreement with the experimental ionization
potential of H$_2$ in \textit{ab initio} DMC calculations it was
necessary to treat both the electrons and the protons as quantum
particles.  Even for heavier atoms than hydrogen, it is unreasonable
to expect quantitative agreement with experiment in the absence of
vibrational corrections.

\subsubsection{O$_2$ dimer\label{subsub:o2_dimer}}

We have performed static-nucleus SJ-DMC ionization-potential
calculations for the O$_2$ molecule, similar to the calculations
described in Sec.\ \ref{subsub:h2_dimer}. Our results are shown in Table
\ref{table:o2_h2_gaps}.

The triplet ground-state ${}^3\Sigma^-_{\rm g}$ electronic configuration was
used to obtain the results given in Table \ref{table:o2_h2_gaps}, with a
geometry obtained from structural relaxation of the triplet state in
spin-polarized DFT, and with explicitly spin-polarized single-particle orbitals
populating the single Slater determinant of orbitals in the trial wave
function.\footnote{We have calculated the energy of the triplet state with and
without the use of spin-polarized DFT orbitals, finding that the spin-polarized
orbitals provide a DMC total energy which is lower, but by a statistically
insignificant amount [0.016(16) eV]\@. We have given results with the
spin-polarized orbitals, owing to the physically reasonable nature of their
use.} However, we have also evaluated the ${}^{1}\Delta_{\rm g}$
singlet-state energy, evaluated with a geometry obtained from structural
relaxation of the singlet state in DFT, finding that it is higher by 1.62(2)
eV than the triplet ground-state energy.  This is rather higher than the
experimental splitting between these two spin configurations of 0.9773
eV\@.\cite{schweitzer2003}

There is an important way in which the single-determinant wave function
we have thus far used to describe the singlet state of O$_2$ might be
inadequate. The singlet state is degenerate at the single-particle level, and
one could in principle find a significantly better singlet wave function by
inclusion of all symmetry-allowed determinants in the subspace of these
degenerate states: at the single-determinant level, the DMC energy of the
singlet state is essentially arbitrary. We have performed multideterminant DMC
calculations for the singlet state, forming a few-determinant expansion with
spin-unpolarized DFT orbitals populating the Slater part of the trial wave
function, and find that the multideterminant singlet ground state energy is
lower in energy by 1.37(2) eV with respect to the single-determinant singlet
state. The DMC
singlet-triplet splitting of O$_2$ is then 0.20(3) eV, which is
significantly lower than the previously quoted experimental value of 0.9773
eV\@.\cite{schweitzer2003}

The underestimate of the singlet-triplet splitting reflects the fact
that the singlet trial wave function has more variational freedom via
the use of multiple (degenerate) determinants.  We could easily
improve the triplet wave function by forming a multideterminant expansion
using nondegenerate determinants.  However, this illustrates a general
difficulty with the use of multideterminant wave functions in QMC
calculations of energy differences.  Most QMC calculations rely on a
cancellation of fixed-node errors and in general it is difficult to
provide multideterminant wave functions of equivalent accuracy for two
different systems.

\subsubsection{Nondimer molecules\label{subsub:non_dimers}}

\begin{figure*}[ht!]
  \centering
  \includegraphics[width=\linewidth]{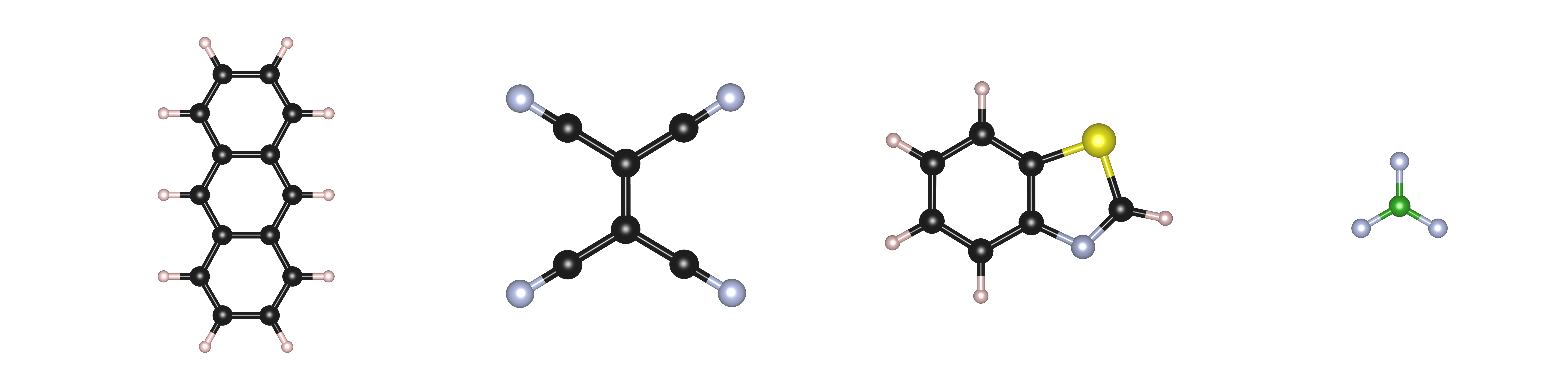}
  \caption{(Color online) Nondimer molecules whose energy gaps we have
    calculated. From left to right: anthracene (C$_{14}$H$_{10}$),
    tetracyanoethylene (C$_6$N$_4$), benzothiazole (C$_7$H$_5$NS), and
    boron trifluoride (BF$_3$).}
\label{fig:molecule_images}
\end{figure*}

The aromatic compounds anthracene (C$_{14}$H$_{10}$) and benzothiazole
(C$_{7}$H$_{5}$NS) are known to possess sizeable first ionization
potentials, as is boron trifluoride (BF$_3$).  Tetracyanoethylene
(C$_6$N$_4$), on the other hand, is a strong Lewis acid, with a large
electron affinity. With this in mind, we have calculated the
ionization potentials and, where positive, the electron affinities of
these molecules using SJ-DMC, with and without the effects of
structural relaxation in the excited state at the DFT level. Our
results for the first three of these molecules are displayed in Table
\ref{table:non_dimer_gaps}. The structures of the molecules we have
studied are shown in Fig.\ \ref{fig:molecule_images}.

\begin{squeezetable}
\begin{table*}[ht!]
\centering
\caption{SJ-DMC ionization potentials and electron affinities of
  various nondimer molecules. Calculations employing relaxed
  excited-state geometries are designated with ``(ER).'' Adiabatic
  gaps are given the subscript ``A'' and vertical gaps the subscript
  ``V\@.''}\label{table:non_dimer_gaps}
\begin{ruledtabular}
\begin{tabular}{l*{6}{c}|*{6}{c}}
\multirow{2}{*}{Molecule} & \multicolumn{6}{c|}{Ionization potential (eV)} &
\multicolumn{6}{c}{Electron affinity
(eV)} \\
  & \mc{SJ-DMC} & \mc{SJ-DMC (ER)} & \mc{$GW$} & \mc{TDDFT} & \mc{CCSD(T)} &
  \multicolumn{1}{c|}{Expt.} & \mc{SJ-DMC} & \mc{SJ-DMC (ER)} & \mc{$GW$} &
  \mc{TDDFT} & \mc{CCSD(T)} & \mc{Expt.} \\ \hline
C$_{14}$H$_{10}$
&7.35(3)&7.31(3)&7.06\cite{Blase2011}&7.02$_\text{A}$\cite{Malloci2011}&
7.52\cite{knight2016accurate}
&7.439(6)$_{\text{A}}$\cite{hager1988} & 0.33(3)& 0.45(3)&0.32\cite{Blase2011}
& 0.53$_{\text{A}}$\cite{Malloci2011}&
0.33\cite{knight2016accurate}&  0.530(5)$_{\text{A}}$\cite{schiedt1997}\\
 & & & &7.09$_{\text{V}}$\cite{Malloci2011}& & & & &
 &0.43$_{\text{V}}$\cite{Malloci2011} & & \\
C$_{7}$H$_{5}$NS &8.92(2)&8.80(2)&8.48\cite{Blase2011}& &
&8.72(5)$_{\text{A}}$\cite{eland1969}&--&--&--&--&--&-- \\
BF$_{3}$ &16.226(6)&16.227(6)& & & &
15.96(1)$_{\text{V}}$\cite{Batten1978}&--&--&--&--&--&--
\end{tabular}
\end{ruledtabular}
\end{table*}
\end{squeezetable}

As an example of an excitonic gap in a molecule, we have evaluated the first
singlet and triplet excitation energies of anthracene at the SJ-DMC level. We
find that the singlet excitation energy is 3.07(3) eV, while the corresponding
triplet excitation energy is 2.36(3) eV\@. A recent QMC study obtained a
significantly larger (vertical) singlet VMC excitation energy of 4.193(17) eV
[4.00(4) eV at the DMC level];\cite{Dupuy2015} however, the form of trial wave
function was qualitatively different, and various details of the underlying
geometry-relaxation and orbital-generation calculations differ from what we
have reported here.  Available experimental values for the singlet excitations
are 3.38\cite{biermann1980diels} and 3.433,\cite{baba2009structure} while a
single experiment (on molecules in a solvent) has claimed that the triplet
excitation energy lies in the range 1.84--1.85 eV\@.\cite{padhye1956lowest}
However, comparison is complicated due to the presence of vibrational effects,
which generally differ for singlet and triplet excitations.

For the cases of C$_6$N$_4$ and BF$_3$ we have also performed some
test SJB calculations. We find that the SJB-DMC ionization potential
of BF$_3$ is
16.221(4) eV [the difference from the SJ-DMC value of 16.226(6) eV
  being statistically insignificant], and present our C$_6$N$_4$
results in Table \ref{table:TCNE_gaps}.  Backflow correlations have
little effect on the calculated ionization potentials and electron
affinities. Nor are the calculated energy differences significantly
affected by the reoptimization of excited-state geometries. We
therefore expect that the dominant sources of error in these cases
arise from the use of pseudopotentials and (in comparisons with
experiment) vibrational renormalization.

\begin{table}[ht!]
\centering
\caption{DMC ionization potentials (IPs) and electron affinities (EAs)
  of C$_6$N$_4$ at various levels of QMC theory, compared to
  experiment and other methods. Calculations employing relaxed
  excited-state geometries are designated ``(ER),'' and those
  employing reoptimized backflow functions ``(R).'' Adiabatic gaps are
  given the subscript ``A,'' vertical gaps the subscript
  ``V\@.''}\label{table:TCNE_gaps}
\begin{ruledtabular}
\begin{tabular}{lcc}
Method   & IP (eV) & EA (eV) \\
\hline
SJ-DMC      & 11.87(1) & 3.23(1)  \\
SJ-DMC (ER) & 11.85(1) & 3.25(1)  \\
SJB-DMC     & 11.88(1) & 3.20(1)  \\
SJB-DMC (ER) & 11.86(1) & 3.23(1)  \\
SJB(R)-DMC  & 11.87(1) & --       \\
SJB(R)-DMC (ER) & 11.84(1) & --       \\
\hline
$GW$ & 11.192--12.517\cite{knight2016accurate} &
3.30--$\sim$3.9\cite{ren2015beyond} \\
& & 2.732--3.804\cite{knight2016accurate} \\
\hline
CCSD(T) & 11.99\cite{richard2016} & 3.05\cite{richard2016} \\
\hline
Expt.   & 11.79(5)$_{\text{V}}$\cite{houk1976}    &
3.16(2)$_{\text{A}}$\cite{khuseynov2012} \\
        & 11.765(8)$_{\text{A}}$\cite{knowles1974} &
\end{tabular}
\end{ruledtabular}
\end{table}

\subsection{Three-dimensional solids\label{sub:solids}}

\subsubsection{Diamond Si\label{subsub:diamond_si}}

\begin{figure}[ht!]
\begin{center}
\includegraphics[clip,width=\linewidth]{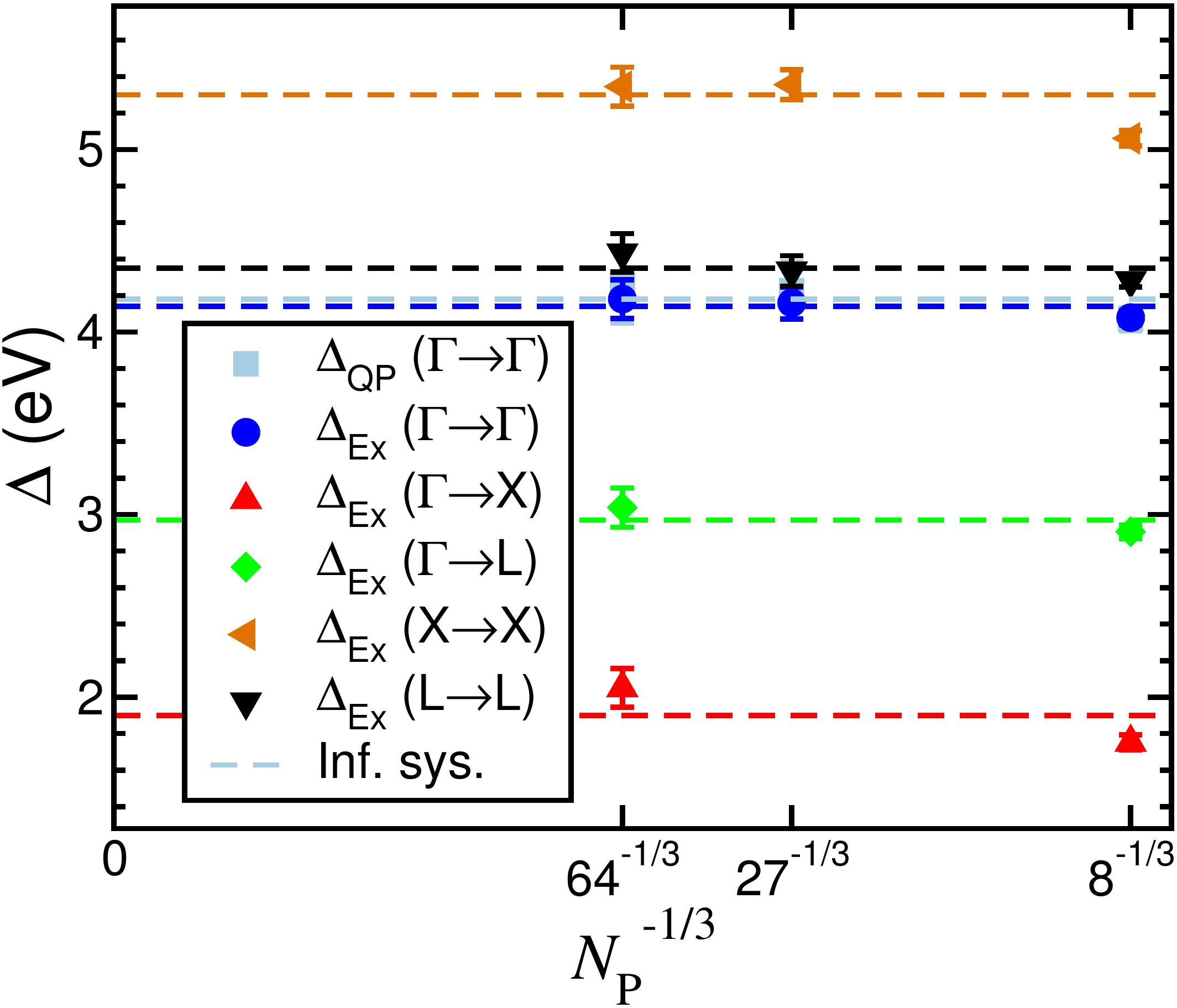}
\caption{(Color online) Finite-size errors in uncorrected SJ-DMC
  quasiparticle and excitonic gaps $\qpg$ and $\exg$ of Si as a
  function of the number of primitive cells $N_{\rm P}$ in the
  supercell.  The dashed lines show the infinite-system gaps estimated
  by subtracting the supercell Madelung constant from the gaps
  obtained in finite cells and averaging over the different
  cells.  \label{fig:si_gaps_vs_NP}}
\end{center}
\end{figure}

Silicon in the diamond structure is an indirect-band-gap semiconductor with a
valence-band maximum at the $\Gamma$ point ($\Gamma_{\rm v}$) in the FCC
Brillouin zone and a conduction-band minimum at around 85\% of the distance
along the line $\overline{\Gamma {\rm X}}$. Extensively studied over the past
few decades by experimentalists and theorists alike, Si provides an ideal
test-bed on which to benchmark QMC band-gap results. To this end, we have
calculated the excitonic gaps of Si between various high-symmetry points in the
Brillouin zone.  Specifically, we have considered promotions from $\Gamma_{\rm
v} \rightarrow \Gamma_{\rm c}$, $\Gamma_{\rm v} \rightarrow {\rm L}_{\rm c}$,
$\Gamma_{\rm v} \rightarrow {\rm X}_{\rm c}$, ${\rm L}_{\rm v} \rightarrow {\rm
L}_{\rm c}$, and ${\rm X}_{\rm v} \rightarrow {\rm X}_{\rm c}$. Calculations of
the $\Gamma_{\rm v} \rightarrow {\rm L}_{\rm c}$ and $\Gamma_{\rm v}
\rightarrow {\rm X}_{\rm c}$ excitonic gaps are forbidden in the
$3\times3\times3$ supercell, where no choice of supercell reciprocal lattice
vector ${\bf k}_{\text{s}}$ can ensure that both ${\rm L}$ and ${\rm X}$ appear
simultaneously with $\Gamma$ in the $3\times3\times3$ grid of $\textbf k$
points used to generate our single-particle orbitals. In order to address the
issue of finite-size effects in our energy gaps, we have used simulation
supercells comprised of $2\times2\times2$, $3\times3\times3$, and
$4\times4\times4$ arrays of primitive cells, and averaged the
finite-size-corrected SJ-DMC results.  The exciton binding energy of Si is very
weak [15.01(6) meV\cite{green2013}], and the exciton Bohr radius is much larger
than the simulation cells available to QMC calculations.  We therefore expect
the excitonic and quasiparticle gaps to be very similar and to show the same
finite-size scaling.  Our energy-gap results are given in Table
\ref{table:si_gaps} and Fig.\ \ref{fig:si_gaps_vs_NP}.

\begin{table*}[!]
\centering
\caption{Uncorrected quasiparticle and excitonic energy gaps $\qpg$ and
  $\exg$ of Si evaluated in SJ-DMC for different simulation
  supercells.}\label{table:si_gaps}
\begin{ruledtabular}
\begin{tabular}{l*{4}{d{1.5}}}
\multirow{2}{*}{Excitation} & \multicolumn{4}{c}{SJ-DMC gap (eV)}\\&
\mc{$2 \times 2 \times 2$ supercell} & \mc{$3 \times 3 \times 3$ supercell}
& \mc{$4 \times 4 \times 4$ supercell} & \mc{FS corr.\ and av.} \\ \hline
$\qpg(\Gamma_{\rm v} \rightarrow \Gamma_{\rm c})$  &
3.56(6)&3.9(2) &4.0(2)& 4.18(6)\\
$\exg(\Gamma_{\rm v} \rightarrow \Gamma_{\rm c})$  &
3.57(4)&3.82(9)&3.9(1)& 4.14(3)\\
$\exg(\Gamma_{\rm v} \rightarrow {\rm X}_{\rm c})$ &
1.24(4)&\mc{--}&1.8(1)& 1.9(1)\\
$\exg(\Gamma_{\rm v} \rightarrow {\rm L}_{\rm c})$ &
2.39(4)&\mc{--}&2.8(1)& 2.97(7)\\
$\exg({\rm X}_{\rm v} \rightarrow {\rm X}_{\rm c})$&
4.55(4)&5.01(8)&5.1(1)& 5.3(1)\\
$\exg({\rm L}_{\rm v} \rightarrow {\rm L}_{\rm c})$&
3.77(4)&4.00(8)&4.2(1)& 4.35(4)
\end{tabular}
\end{ruledtabular}
\end{table*}

\begin{figure}[ht!]
  \centering
  \includegraphics[width=\linewidth]{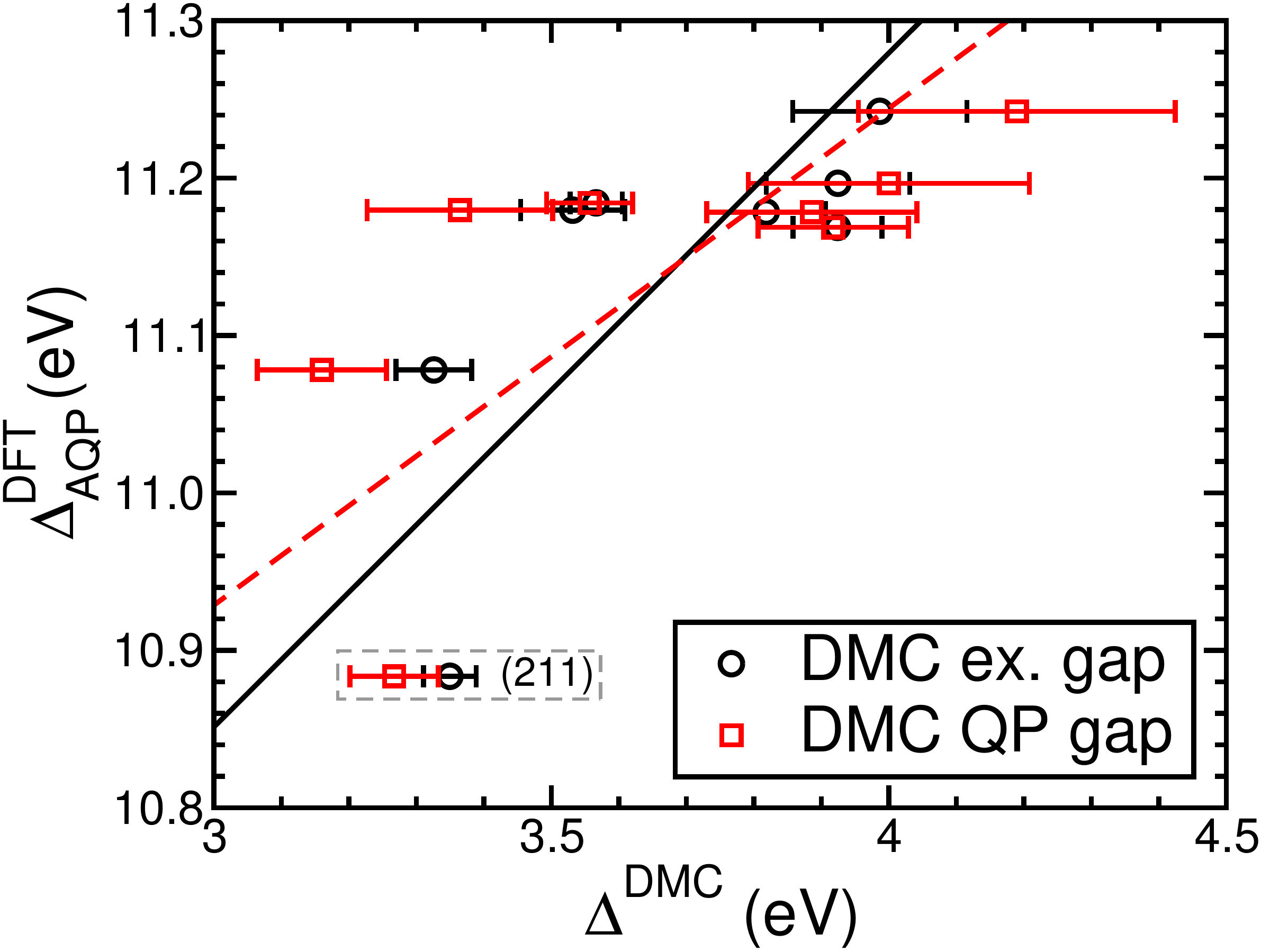}
  \caption{(Color online) Uncorrected SJ-DMC quasiparticle and
    excitonic energy gaps of Si at $\Gamma$, plotted against DFT analog
    ``quasiparticle'' (AQP) gaps, obtained using the defect
    formation energies for positive and negative charged defects.  The
    results were obtained in different sizes and shapes of periodic
    cell.  The straight lines are linear fits of SJ-DMC gap against
    DFT AQP gaps.}
\label{fig:dft_vs_dmc_gaps}
\end{figure}

As a further test of our method and our treatment of finite-size effects, we
have calculated the quasiparticle energy gap at the $\Gamma$ point.  We have
also calculated excitonic and quasiparticle gaps at the $\Gamma$ point in
various differently shaped (noncubic, but diagonal)
supercells.\footnote{Specifically, noncubic cells comprised of: $2\times
1\times 1$, $3\times 1\times 1$, $2\times 2\times 1$, $3\times 2\times 1$, and
$3\times 3\times 1$ arrays of primitive cells.} The results of this
investigation are given in Table \ref{table:si_shapes_gaps}, showing the
quasirandom variation with cell shape. We have found that the finite-size
effects that exist in our SJ-DMC energy-gap data correlate with those obtained
from DFT calculations wherein charged defects have been introduced.
Specifically, we have calculated the DFT total energies of supercells of
intrinsic Si, Si with one P substitution, and Si with one Al substitution, with
the total number of electrons fixed to that of the intrinsic Si calculation.
This mimics the introduction of two point charges, and a DFT analog
quasiparticle gap can be defined as \begin{equation}
\Delta_{\text{AQP}}^{\text{DFT}} = E_{\text{P}} + E_{\text{A}} -
2E_{\text{Si}}, \end{equation} where $E_{\text{X}}$ is the energy of the Si
system with one substitution of atom type X. Our analog DFT energies have been
obtained with a fixed (dense) ${\bf k}$ point sampling, and with ultrasoft
pseudopotentials generated on-the-fly in \textsc{castep}.\footnote{Versions of
\textsc{castep} before 17.2 were subject to a bug which led to incorrect total
energies in charged calculations. We have worked with version 17.2, avoiding
the undesirable behavior. We also note that tests with earlier versions of
\textsc{castep} indicate that the errors in individual total energies reported
by \textsc{castep} do \textit{not} cancel when one calculates a defect
formation energy. We thank S.\ Murphy for drawing our attention to this issue.}
A plot of $\Delta_{\text{AQP}}^{\text{DFT}}$ against $\qpg$ obtained from
SJ-DMC simulations is given in Fig.\ \ref{fig:dft_vs_dmc_gaps}. The correlation
is statistically significant with or without the inclusion of the data point
corresponding to the smallest cell size.  This directly confirms that
finite-size errors in QMC gap calculations are analogous to those in DFT
defect-formation-energy calculations.

\begin{table}[ht!]
\centering
\caption{Finite-size-corrected SJ-DMC vertical quasiparticle gaps
  $\qpg$ and SJ-DMC vertical excitonic gaps $\exg$ at the $\Gamma$
  point in Si for various noncubic supercells. After correction, the
  $\qpg$ and $\exg$ data sets have lower variances by a factor of more
  than two.}\label{table:si_shapes_gaps}
\begin{ruledtabular}
\begin{tabular}{lc *{2}{d{1.5}}}
\multirow{2}{*}{Supercell} & Madelung const. &
\multicolumn{2}{c}{SJ-DMC gap (eV)}\\
                    &(eV) & \mc{$\qpg$} & \mc{$\exg$} \\ \hline
2$\times$1$\times$1 & $-0.7364$ & 4.00(7) &  4.08(4)  \\
2$\times$2$\times$1 & $-0.6009$ & 3.8(1)  &  3.93(6)  \\
3$\times$1$\times$1 & $-0.4116$ & 4.3(1)  &  4.34(7)  \\
3$\times$2$\times$1 & $-0.4370$ & 3.8(1)  &  3.97(8)  \\
3$\times$3$\times$1 & $-0.3342$ & 4.5(2)  &  4.3(1)
\end{tabular}
\end{ruledtabular}
\end{table}

\begin{figure}[ht!]
  \centering
  \includegraphics[width=\linewidth]{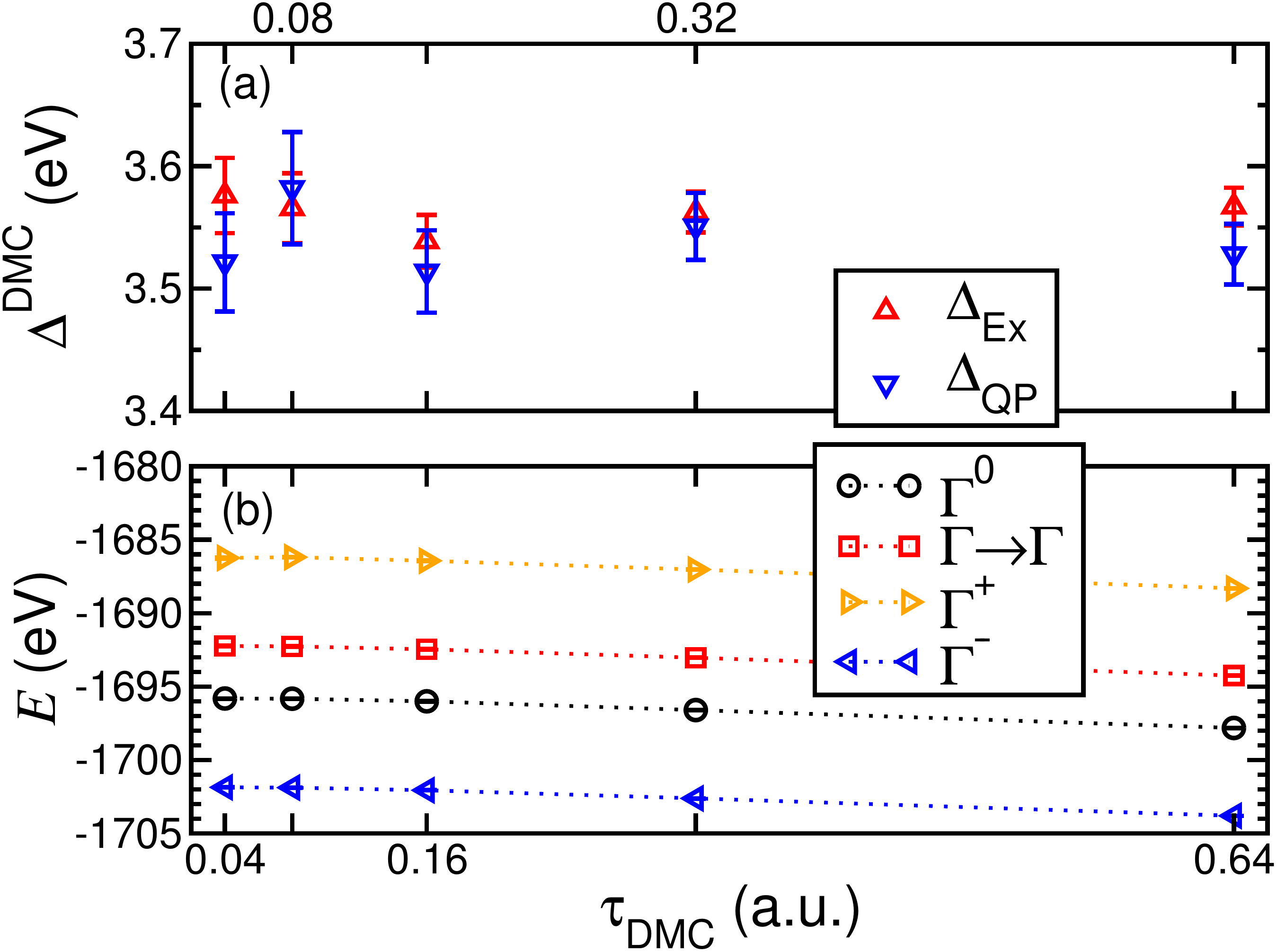}
  \caption{(Color online) Time-step bias in (a) SJ-DMC energy gaps and
    (b) SJ-DMC total energies for ground ($\Gamma^{0}$), excitonic
    ($\Gamma \rightarrow \Gamma$), cationic ($\Gamma^{-}$), and
    anionic ($\Gamma^{+}$) states of Si. All calculations have
    been performed in a 2$\times$2$\times$2 supercell with a
    target population of 256 walkers. The Madelung correction is not included
    (and would only offset the gaps by a constant).}
\label{fig:ts_tests}
\end{figure}

All of our DMC calculations for this system have employed time steps
of 0.01 and 0.04 a.u., except for our tests in noncubic cells, and our
SJB tests, which employed larger time steps of 0.04 a.u.\ and 0.16 a.u
(with a computational speed-up factor of four).
However, we have observed in tests that, in conjunction
with the T-move scheme,\cite{Casula2006} it is possible to use far
larger time steps in SJ-DMC gap calculations. The results of these
tests are displayed in Fig.\ \ref{fig:ts_tests}. While time-step bias
in total energies is significant at larger DMC time steps (of order a
few eV), this bias cancels almost entirely in both excitonic and
quasiparticle energy gaps at fixed system size and DMC population size. We
expect that the use of even larger DMC time steps in other systems
could allow for computational savings of at least an order of
magnitude.

Our largest family of calculations for Si, those in the $4\times
4\times 4$ supercell, required around 1.7 million core hours to
complete. Had we opted to employ time steps of 0.04 and 0.16 a.u.,
which are still conservative choices in light of the information
presented in Fig.\ \ref{fig:ts_tests}, we would have required 0.5
million core hours of computer time.

To address the impact of fixed-node errors in our calculated energy gaps, we
have carried out tests including backflow correlations in our trial wave
functions. We find that the inclusion of backflow lowers the
(Madelung-corrected) DMC excitonic and quasiparticle gaps in a $2\times 2\times
2$ supercell of Si from the SJ-DMC values of 4.08(4) and 4.07(6) eV to the
SJB-DMC values of 3.95(1) and 3.95(3) eV, respectively.  This is an $O(0.1\
\text{eV})$ effect, which we expect to affect our results at larger system
sizes to at least a similar extent.  However, to explicitly verify this for the
larger cells would incur significant further computational expense.
Furthermore, we have considered the impact of reoptimization of backflow
functions in excited states. We find that in the case of the $\Gamma_{\rm v}
\rightarrow \Gamma_{\rm c}$ quasiparticle gap, this reoptimization lowers the
(Madelung-corrected) SJB-DMC gap even further, from $3.95(3)$ to $3.77(3)$ eV\@ in a
$2\times 2\times 2$ cell. For the $\Gamma_{\rm v} \rightarrow 0.85{\rm X}_{\rm
c}$ quasiparticle gap in a $2\times 2\times 2$ supercell, reoptimization lowers
the (Madelung-corrected) SJB-DMC gap from $1.66(2)$ to $1.46(1)$ eV\@. In summary, the
use of SJ trial wave functions leads to positive fixed-node errors in energy
gaps of at least 0.2 eV for Si.  In a material with a negligible exciton
binding energy such as Si, one may choose to calculate ``the gap'' as either an
excitonic gap or a quasiparticle gap; both exhibit the same finite-size errors.
The quasiparticle gap allows the safe reoptimization of backflow functions when
electrons are added to or removed from a simulation supercell, and furthermore
the quasiparticle gap can be calculated between any pair of wavevectors in any
supercell. On the other hand, the excitonic gap requires just two QMC
calculations to be performed in each simulation cell, rather than three or four
for the quasiparticle gap.

A further potential source of fixed-node error at the $\Gamma$ point
arises from the three-fold degeneracy of the
light-hole, heavy-hole, and ``spin-orbit split-off'' bands. Here, a DFT
code will output three arbitrary linear combinations of the
single-particle orbitals in question. To investigate the possible
consequences of this, we have performed SJ-VMC test calculations with
trial wave functions formed from three determinants including each of
the three degenerate single-particle states at $\Gamma$. We find that
the formation of a few-determinant expansion has, in this case, no
statistically significant effect on the resultant quasiparticle band
energy. We have further investigated the potential impact of
degeneracy by repeating these calculations on a grid with ${\bf
  k}_{\text{s}} \ne {\bf 0}$. Here, the $\Gamma$ point is not
explicitly sampled, but instead the grid is centered on a wave vector
of very small magnitude, ${\bf k}_{\text{s}}=(\epsilon, \epsilon',
\epsilon'')$, so as to break the three-fold degeneracy of the orbitals
at $\Gamma$. Here, we again find no change in the resultant
quasiparticle band energy: if all three determinants are included in
the expansion, the total energy of the cationic state at the SJ-VMC
level is $-7.8179(1)$ a.u. The total energy of the single-determinant
state is (again) $-7.8179(1)$ a.u., while the total energies
corresponding to singlet excitations made from the two other (once
degenerate) states are $-7.8177(1)$ and (again) $-7.8177(1)$ a.u. The
differences are statistically insignificant, and we have therefore
eliminated degeneracy as a source of error at the $\Gamma$ point.

Early QMC studies on solids had claimed some success in the evaluation
of band structures and energy gaps. The earliest examples of such
calculations [diamond in Refs.\ \onlinecite{mitasbook,Towler2000},
Si in Ref.\ \onlinecite{Williamson1998}, solid atomic (I2$_1$3)
N in Ref.\ \onlinecite{mitas1994quantum}, and manganese (II)
oxide in Ref.\ \onlinecite{lee2004quantum}] considered direct
calculation of the excitonic gap in small supercells [8 atoms for
diamond, Si, and solid N, 16--20 atoms in manganese (II)
oxide]. Quasiparticle energy gaps were evaluated, if at all, by
means of an addition of an estimate of the exciton binding energy (in
the Mott-Wannier model, for example). SJ trial wave functions were
used exclusively, and no attempts were made to examine explicitly the
nature of finite-size effects in energy gaps themselves, or to explore
fixed-node errors. In common supercell shapes the Madelung constant is
typically \textit{negative}, so that a positive correction to
quasiparticle gaps is required; this would have been partially offset
by fixed-node errors. N.b., the cells used in QMC studies of Si are
small compared with the exciton Bohr radius, so finite-size errors in
the excitonic gap behave the same as finite-size errors in the
quasiparticle gap (see Sec.\ \ref{sub:fs_effects}).

Our QMC quasiparticle gaps in silicon are generally larger than
those obtained from $GW$ calculations. For example, a recent
all-electron $G_0W_0$ calculation determined the $\Gamma \rightarrow
\Gamma$, $\Gamma \rightarrow {\rm X}$, $\Gamma \rightarrow {\rm L}$,
${\rm X} \rightarrow {\rm X}$, and ${\rm L} \rightarrow {\rm L}$
quasiparticle gaps of silicon as 3.07, 0.95, 2.21, 3.46, and 4.09 eV,
respectively.\cite{Ishii2010} A different study determined somewhat
larger (pseudopotential) quasiparticle self-consistent $GW$
quasiparticle gaps from $\Gamma \rightarrow \Gamma$, $X$, and $L$ as
3.54, 1.60, and 2.41 eV, respectively.\cite{Bruneval2006}

\subsubsection{Cubic boron nitride\label{subsub:cbn}}

Cubic BN has the zincblende crystal structure, with
diamond-structure sites alternately occupied by B and N atoms. It is an
insulator with a large and indirect fundamental gap from $\Gamma_{\rm v}
\rightarrow {\rm X}_{\rm c}$. Experimental estimates of the indirect excitonic
gap range from $5.5$--$7.0$ eV,\cite{fomichev1968,chrenko1974} and previous DFT
investigations give a range for the indirect quasiparticle gap from
$4.2$--$8.7$ eV\@.\cite{xu1991,rodriguez1995} Theoretical studies based on
DFT\cite{Cappellini2001} and on the Bethe-Salpeter equation,\cite{Satta2004}
predict that many-body effects in the absorption spectra of cubic BN are
significant, and that a Mott-Wannier exciton formed between the valence and
conduction bands at $\Gamma$, with binding energy around 0.35 eV, should exist
in the bulk material.  We have calculated the excitonic energy gaps of cubic BN
between the same high-symmetry points as for Si, and have also calculated the
quasiparticle gap from $\Gamma_{\rm v} \rightarrow \Gamma_{\rm c}$.  Our
energy-gap results for cubic BN are given in Table \ref{table:cbn_gaps}. We
find that the quasiparticle gap from $\Gamma_{\rm v} \rightarrow \Gamma_{\rm
c}$ is $12.8(2)$ eV, but are unable to resolve a statistically significant
$\Gamma_{\rm v} \rightarrow \Gamma_{\rm c}$ exciton binding energy, because our
SJ-DMC error bars are $\sim 0.2$ eV, compared to the expected exciton binding
of around 0.35 eV\@.  Our value of $7.5(3)$ eV for the indirect excitonic gap
is consistent with the range of experimental estimates.

\begin{table*}[!]
\centering
\caption{Uncorrected quasiparticle and excitonic energy gaps $\qpg$ and $\exg$
  of cubic BN evaluated in SJ-DMC for different simulation
  supercells.}\label{table:cbn_gaps}
\begin{ruledtabular}
\begin{tabular}{l*{4}{d{1.5}}}
\multirow{2}{*}{Excitation} & \multicolumn{4}{c}{SJ-DMC gap (eV)}\\&
\mc{$2\times 2 \times 2$ supercell} & \mc{$3 \times 3 \times 3$ supercell}
& \mc{$4 \times 4 \times 4$ supercell} & \mc{FS corr.\ and av.} \\ \hline
$\exg(\Gamma_{\rm v} \rightarrow \Gamma_{\rm c})$ &
10.45(4)&11.60(9)&12.06(4) &12.9(2)\\
$\qpg(\Gamma_{\rm v} \rightarrow \Gamma_{\rm c})$ &
10.37(5)&11.7(2)&12.00(8)  &12.8(2)\\
$\exg(\Gamma_{\rm v} \rightarrow {\rm X}_{\rm c})$&
5.12(4) &\mc{--}&6.76(5)   &7.5(3)\\
$\exg(\Gamma_{\rm v} \rightarrow {\rm L}_{\rm c})$&
11.67(4)&\mc{--}&13.16(4)  &14.0(2)\\
$\exg({\rm X}_{\rm v} \rightarrow {\rm X}_{\rm c})$&
10.77(4)&11.85(8)&12.50(5)&13.2(2)\\
$\exg({\rm L}_{\rm v} \rightarrow {\rm L}_{\rm c})$&
13.60(4)&14.81(8)&15.37(5)&16.1(2)
\end{tabular}
\end{ruledtabular}
\end{table*}

\subsubsection{$\alpha$-quartz: SiO$_2$\label{subsub:sio2}}

The $\alpha$-quartz polymorph of SiO$_2$ is the most thermodynamically stable
at ambient conditions, and hence common.  Recent quasiparticle self-consistent
$GW$ (QSGW) calculations\cite{kresse2012} corroborate earlier theoretical
claims\cite{chang2000} that the system hosts a very-well-bound exciton formed
at the $\Gamma$ point in the Brillouin zone. The exciton binding energy
obtained in Ref.\ \onlinecite{kresse2012} is 1.2 eV, compared with 1.7 eV in
Ref.\ \onlinecite{chang2000}. Experiment finds that the exciton binding is
around 1 eV\@.\cite{philipp1966}  We have calculated the quasiparticle and
excitonic gaps from $\Gamma_{\rm v} \rightarrow \Gamma_{\rm c}$, in
$1\times1\times1$ and $2\times2\times2$ supercells in an attempt to explore
this phenomenon. The crystal structure of $\alpha$-quartz makes the study of
larger supercells prohibitively expensive (the unit cell consists of three Si
atoms and six O atoms, or 48 electrons when using Trail-Needs pseudopotentials
to describe core electronic states). We find that the SJ-DMC quasiparticle and
excitonic gaps of $\alpha$-SiO$_2$ are 11.4(2) eV and 11.51(7) eV,
respectively. We are hence unable to extract a statistically significant
exciton binding in $\alpha$-SiO$_2$, perhaps due to the limited sizes of
simulation cell that we can study in this case.

\subsection{Two-dimensional phosphorene\label{subsub:phosphorene}}

\begin{figure}[ht!]
  \centering
  \resizebox{0.8\columnwidth}{!}{\includegraphics[clip]{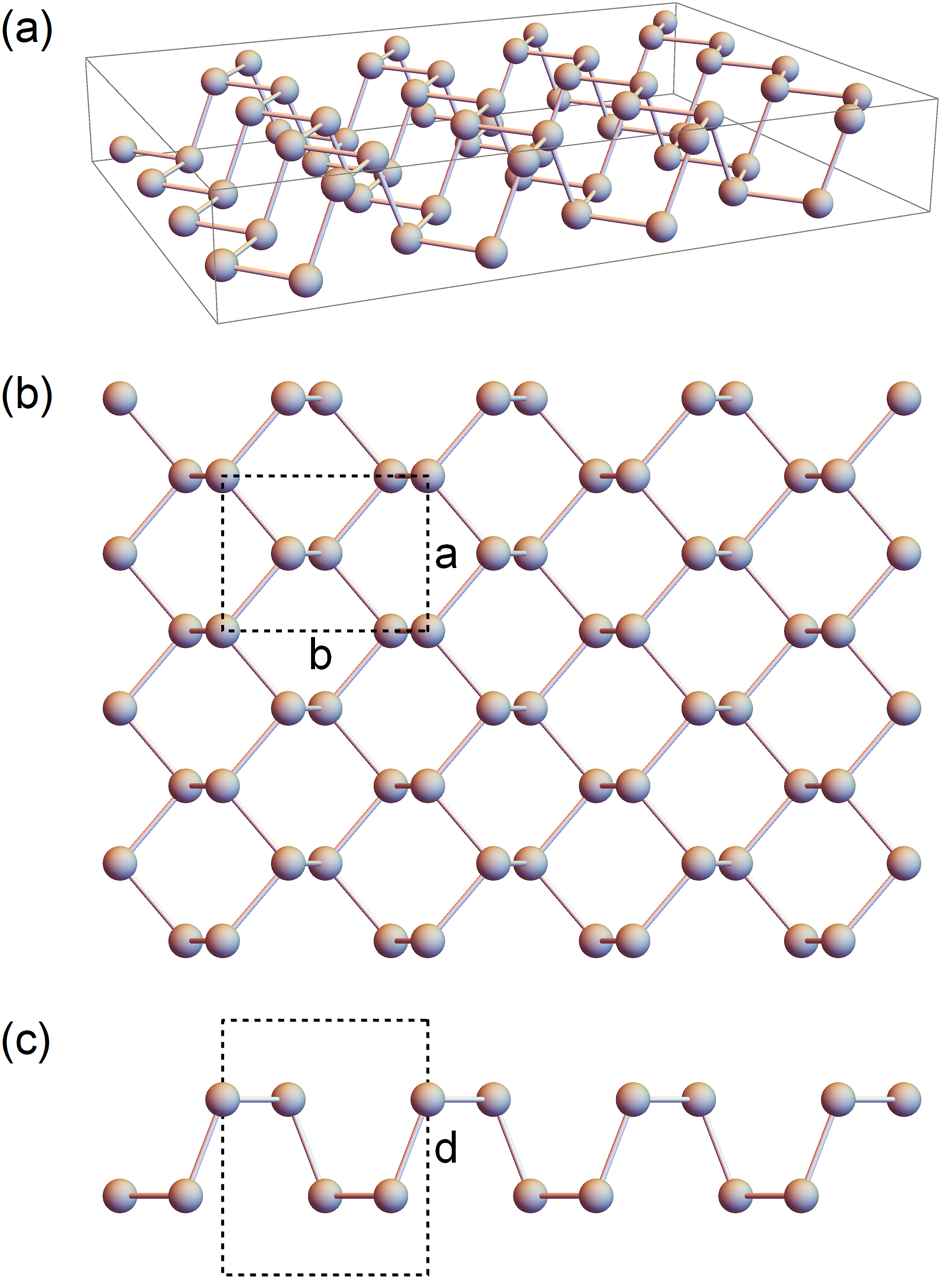}}
  \caption{(Color online) Geometry of a phosphorene layer: (a)
    tilted view, (b) top view, and (c) front view.\label{fig:phos_geom}}
\end{figure}

Phosphorene (monolayer black phosphorus) is a 2D material that
exhibits a large exciton binding according to $GW$-BSE
calculations,\cite{Tran2014,Tran2015,Choi2015} an effective-mass model
parameterized by DFT,\cite{Seixas2015} and experimental studies of
few-layer black phosphorus on a substrate together with an
effective-mass model.\cite{Zhang2018} Phosphorene consists of
phosphorus atoms, four in each unit cell, in a 2D armchair structure
with a rectangular Bravais lattice: see Fig.\ \ref{fig:phos_geom}.  We
used DFT-PBE to obtain a relaxed geometry with lattice parameters $a =
3.31$ {\AA} and $b = 4.56$ {\AA}\@.  As a 2D material, the screened
interaction between charge carriers is of Keldysh form, and care is
required in the treatment of finite-size effects.  The electron and
hole effective masses $m_{\rm e}^\ast=0.44\ m_0$ and $m_{\rm
  h}^\ast=0.98\ m_0$ may be roughly estimated as geometrical means of
the masses in the zig-zag and armchair directions;\cite{Qiao2014} the vacuum
in-plane susceptibility parameter is estimated to be $r_\ast=24.24$
{\AA}\@.\cite{Seixas2015} The physical size of the exciton in the
effective-mass approximation is therefore $r_0=\sqrt{r_*/(2\mu)}=4.6$
{\AA} for free-standing phosphorene in vacuum.

Due to phosphorene's anisotropic nature we studied simulation
supercells comprised of $2 \times 2$, $3 \times 2$, $4 \times 3$, $5
\times 4$, and $7 \times 5$ primitive cells. Each supercell was chosen
to be as square as possible, maximizing the nearest-image distance in
the space of diagonal supercells.  The radii of the largest spheres
that can be inscribed in the Wigner-Seitz cells of the simulation
supercells are 3.3, 4.6, 6.6, 8.3, 11.4 {\AA}, respectively.  Thus we
are in the regime in which the Keldysh interaction must be used to
evaluate the Madelung correction to the quasiparticle gap, with the
correction being roughly independent of system size, at least for the
smaller cells. We exclude the $2 \times 2$ supercell from our
extrapolation of the excitonic gap to the thermodynamic limit, since
it is too small to contain the exciton.  Residual finite-size errors
in the Madelung-corrected quasiparticle gap and in the excitonic gap
are expected to scale as $1/L^2$, i.e.\ as $1/N_{\rm P}$, where
$N_{\rm P}$ is the number of primitive cells, over our range of
supercell sizes (this would cross over to $1/L^3$ behavior if the
supercell size exceeded $r_*$).  We have also studied one nondiagonal
supercell containing six primitive cells, which has a slightly larger
Wigner-Seitz cell radius (4.9 {\AA}) than the $3 \times 2$ supercell.
We find that the energy gaps in the nondiagonal cell differ from those
obtained in the $3 \times 2$ supercell by amounts which are not
statistically significant.

\begin{figure}[ht!]
  \centering
  \resizebox{0.95\columnwidth}{!}{\includegraphics[clip]{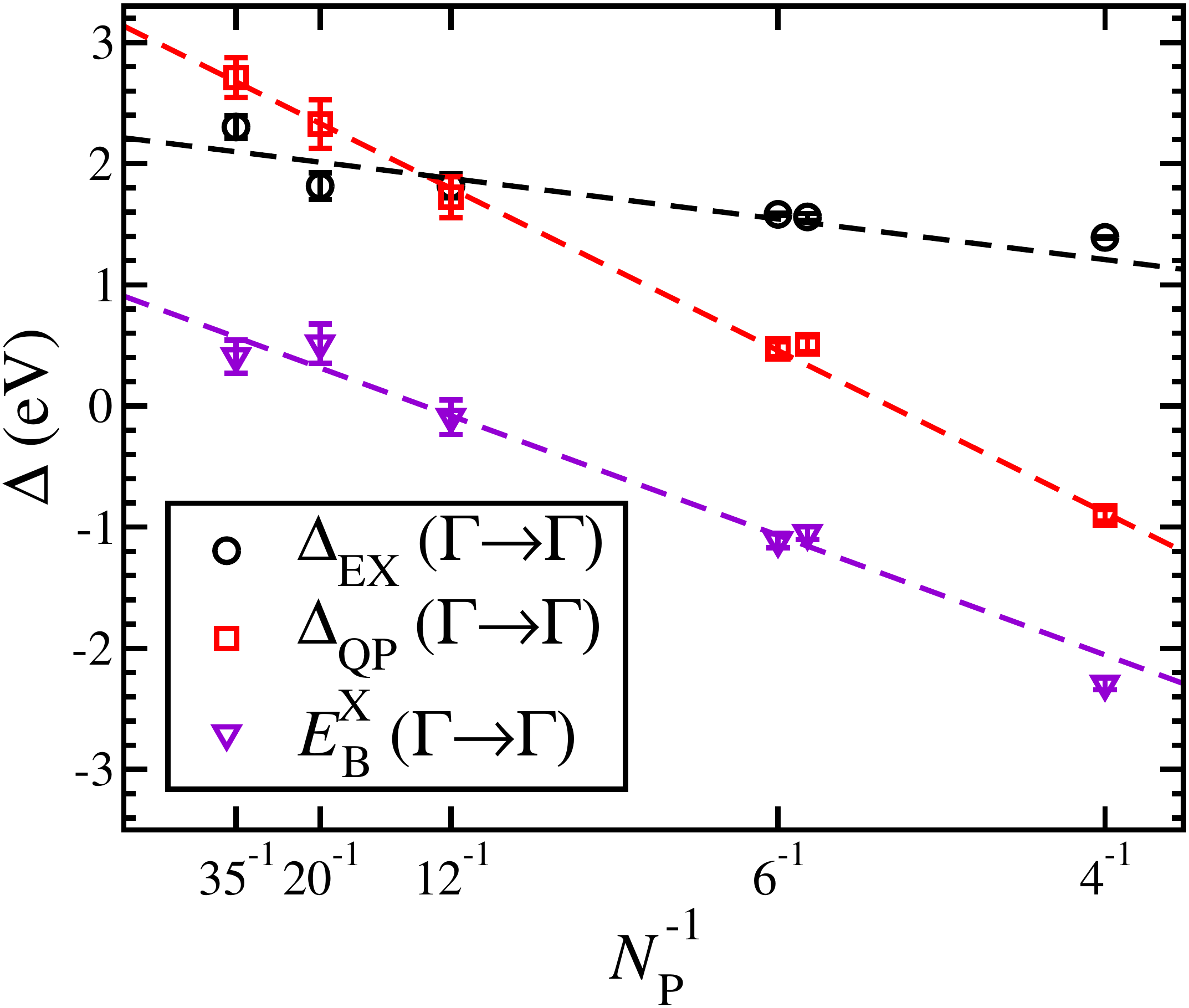}}
  \caption{(Color online) DMC quasiparticle gaps $\Delta_{\text{QP}}$,
    excitonic gaps $\Delta_{\text{Ex}}$, and exciton binding energies
    $E_{\text{B}}^{\text{X}}$ at $\Gamma$ against the inverse
    of the number $N_{\rm P}$ of primitive cells in the supercell
    for a free-standing phosphorene monolayer. The Keldysh Madelung constant
    correction has been applied to the quasiparticle gaps; no finite-size
    correction has been applied to the excitonic gaps. The nondiagonal
    supercell results (filled symbols) have been slightly shifted relative to
    the $3\times2$ supercell result for readability.
\label{fig:phosphorene_gaps}}
\end{figure}

Our results for the excitonic gap $\Delta_{\text{Ex}}$, the
quasiparticle gap $\Delta_{\text{QP}}$, and the exciton binding energy
$E_{\text{B}}^{\text{X}}$ are shown in
Fig.\ \ref{fig:phosphorene_gaps}.  $\Delta_{\text{QP}}$ and
$E_{\text{B}}^{\text{X}}$ have been corrected with the Keldysh
Madelung constant, which was evaluated using the same procedure as the
Madelung constant of the 2D Coulomb interaction, but with the
reciprocal-space interaction being $2\pi/[q(1+r_*q)]$ rather than
$2\pi/q$.\cite{Ewald1921} We then extrapolate the excitonic gap and
Madelung-corrected quasiparticle gap to the thermodynamic limit
assuming the error scales as $1/L^2$ (i.e., we neglect the effects of
the crossover to $1/L^3$ scaling at $L \sim r_*$).

The resulting energy gaps are slightly larger than
previous
estimates\cite{Tran2014,Tran2015,Choi2015,Seixas2015,Zhang2018} for a
free-standing phosphorene monolayer, but our exciton binding energy is
consistent with these estimates, as shown in Table
\ref{table:phosphorene_results}.

\begin{table*}[ht!]
  \caption{Comparison of the SJ-DMC energy gaps and exciton binding of
    monolayer phosphorene with results available in the literature for a
    free-standing monolayer and a monolayer on an SiO$_2$
    substrate.
\label{table:phosphorene_results}}
  \begin{tabular}{lcccc}
    \hline \hline
    Environment & Method & $\Delta_{\text{QP}}$ (eV) &
    $\Delta_{\text{Ex}}$ (eV) & $E_{\text{B}}^{\text{X}}$ (eV)\\
    \hline

    Vacuum & (SJ-DMC, linear extrapolation in $N_{\rm P}^{-1}$) & 3.13(4)
    & 2.2(2) & 0.9(1)\\



    Vacuum &
    Effective-mass approx.,\cite{Tran2014,Tran2015,Choi2015,Seixas2015}
    effective-mass approx./experiment\cite{Zhang2018} & 2.0--2.26 & 1.2--1.41 &
    0.762--0.85\\

    SiO$_2$ substrate & Theory\cite{CastellanosGomez2014,Chaves2015}
    & 2.15 & 1.77 & 0.38--0.396 \\

    SiO$_2$ substrate & Experiment\cite{Liang2014,Yang2015} & 2.05 & 1.75 &
    0.3 \\
  \hline \hline
  \end{tabular}
\end{table*}

We have explicitly tested the effect of a backflow transformation in
the optimal nondiagonal $N_{\rm P}=6$ supercell of phosphorene,
finding that the inclusion of a backflow transformation (optimized in
the ground state) has no statistically significant effect on the DMC
energy gaps. The SJB-DMC quasiparticle gap is 0.03(9) eV lower in
energy than the SJ-DMC quasiparticle gap, and the SJB-DMC excitonic
gap is 0.04(5) eV lower in energy than the SJ-DMC excitonic gap.

\begin{figure}[ht!]
  \centering
  \resizebox{0.75\columnwidth}{!}{\includegraphics[clip]{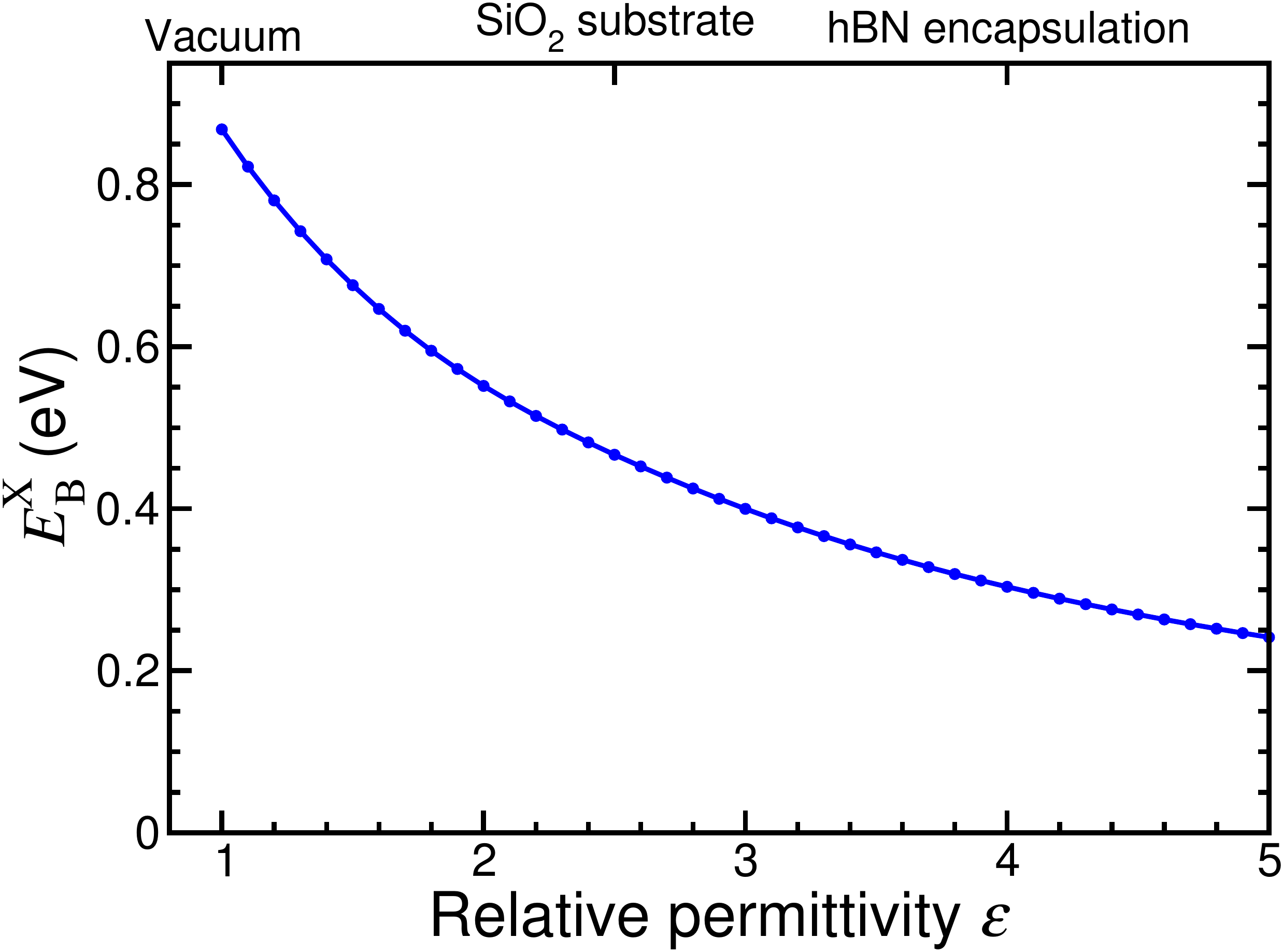}}
  \caption{(Color online) Exciton binding energy in phosphorene
          within the effective-mass approximation with the Keldysh
          interaction between charges as a function of the permittivity
          of the surrounding medium $\epsilon$. Results
          were obtained using the fitting formula from
          Ref.\ \onlinecite{Mostaani2017}. \label{fig:phosphorene_Keldysh}}
\end{figure}

Comparison with experiment is complicated by the fact that the exciton
binding energy is strongly dependent on the dielectric environment of
the monolayer sample.  For example, available
theoretical\cite{CastellanosGomez2014,Chaves2015} and
experimental\cite{Liang2014,Yang2015} results for phosphorene on a
SiO$_2$ substrate show a decrease in the exciton binding and a larger
excitonic gap, as compared to vacuum results.  Using the exciton
fitting function developed in Ref.\ \onlinecite{Mostaani2017} and
phosphorene parameters available in the
literature,\cite{Seixas2015,Qiao2014} we show the dependence of the
exciton binding energy on the dielectric medium surrounding the
monolayer in Fig.\ \ref{fig:phosphorene_Keldysh}.

We attempted an \textit{ab initio} calculation of the biexciton
binding energy in monolayer phosphorene.  The biexciton binding energy
is $E_{\text{B}}^{\text{XX}} = 2 E_N^{+} - E_N - E_N^{++}$, where
$E_N^{+ +}$ is the energy of a doubly promoted $N$-electron system.
$E_{\text{B}}^{\text{XX}}$ was calculated to be $- 29 (10)$ meV and
$16 (13)$ meV in $2 \times 2$ and $3 \times 2$ supercells,
respectively. The former of these cells is too small to describe the
exciton, let alone the biexciton, and the latter cell is too small for
the biexciton.  Unfortunately it was infeasibly expensive for us to
look at larger cells with the required precision.

A very recent QMC study of phosphorene has explored the use of
``hard-wall'' boundary conditions for the evaluation of energy gaps,
by studying hydrogen-terminated molecular flakes of
phosphorene.\cite{Frank2018} In this case, the dominant finite-size
effect appears as an $O(L^{-2})$ confinement effect in the kinetic
energy of the added or removed charge rather than the slowly decaying
image-interaction effect that occurs in a periodic supercell.

For 3D crystals, it is relatively straightforward to remove the
$O(L^{-1})$ finite-size error in the quasiparticle gap under periodic
boundary conditions by using the Madelung correction.  The use of
finite clusters to approximate the bulk introduces other nonsystematic
finite-size errors, such as edge-termination effects.  Indeed, the
nature of the electronic states involved in the excitation are not
necessarily even qualitatively similar to the relevant electronic
states in the infinite system.  For example, the lowest unoccupied
molecular orbital in a diamondoid molecule is a delocalized surface
state that does not correspond to the bulk diamond conduction-band
minimum,\cite{drummond2005electron} and were one to attempt to
calculate the band gap of bulk diamond by consideration of larger and
larger diamondoids one would have to address this difficulty.

For 2D materials, however, hard-wall boundary conditions provide a
relatively attractive method for \textit{ab initio} calculations of
quasiparticle gaps and the energies of charged excitations. As shown
here, calculations in periodic supercells smaller than $r_*$ are
absolutely dependent on a Madelung correction evaluated using the
Keldysh interaction; since this is roughly constant in cells with
$L<r_*$, it is not possible even in principle to extrapolate gaps to
the thermodynamic limit.  By contrast, gaps obtained in
hydrogen-terminated flakes can be extrapolated to infinite size
without relying on model interactions.  For excitonic gaps the
finite-size errors go as $1/L^2$ under periodic boundary conditions on
supercells with $L<r_*$, and hence can be extrapolated if the crossover
to $1/L^3$ behavior is neglected.  In this case calculations using
periodic boundary conditions maybe preferable, as they are less
affected by surface effects.

We emphasize that the need for large periodic cells to describe
charged quasiparticles in 2D materials is not an artifact of QMC
calculations, but an inevitable consequence of the physics of 2D
materials and the Keldysh interaction, which must affect all attempts
at \textit{ab initio} gap calculations in these materials.  Similar
considerations must arise in calculations of charged defect formation
energies in layered and 2D materials.

\section{Conclusions\label{sec:conclusions}}

We have reviewed the use of QMC methods to calculate energy gaps in
atoms, molecules, and crystals.  Although the quasiparticle gap does
not formally satisfy a variational principle, in practice the
fixed-node error in the quasiparticle gap is overwhelmingly likely to
be positive.  Reoptimization of trial wave functions for systems in
which electrons have been added or removed can be expected to improve
the calculated quasiparticle gaps. For neutral excitations (excitonic
promotions) this is not necessarily the case, as was demonstrated in
Sec.\ \ref{sub:qp_ex_gaps}, and reoptimization can potentially result
in the formation of a pathological excited-state trial nodal surface.
Unless the neutral excitation results in a trial wave function that
transforms as a 1D irrep of the full
symmetry group of the system and the target state is the lowest-energy
eigenstate that transforms as that irrep,
reoptimization of the free parameters in the excited-state wave
function should not be attempted.  Since Jastrow factors do not affect
the nodal surface and hence DMC energy, there is little to be gained
by reoptimizing Jastrow factors in excited states; on the other hand,
reoptimizing backflow functions in states in which electrons have been
added to or removed from the neutral ground state can significantly
improve DMC quasiparticle gaps.

The use of larger-than-typical DMC time steps for excitation calculations
has been shown to be a major source of possible computational savings in DMC
energy-gap calculations. Time-step bias appears to cancel extraordinarily well
in energy gaps. In Si we have made computational savings of a factor of four by
using larger time steps in backflow calculations.

Our calculations employing multideterminant trial wave functions for
Si at the $\Gamma$ point show that, even where bands are exactly
degenerate, it is not necessarily the case that a few-determinant
excited-state wave function comprised of contributions from all
possible combinations of degenerate single-particle orbitals performs
any better than the single-determinant alternative\@.  On the other
hand, such a multideterminant wave function significantly lowers the energy
of the singlet first-excited state of O$_2$.  The need for multideterminant
wave functions appears to be more of an issue in studies of molecules than
crystalline solids.

We have evaluated energy gaps in atomic, molecular, and solid systems
using the VMC and DMC methods with single-determinant SJ and SJB trial
wave functions.  In atomic Ne, where vibrational and finite-size
effects are not present, we have achieved highly accurate ionization
potentials in comparison with experimental data from which
relativistic effects have been removed. The MAE across all of our
SJB-DMC calculated ionization potentials for Ne is 0.34\%,
demonstrating the intrinsic high accuracy achieved by the SJB-DMC
method.

In various molecules, where vibrational effects may be present, but
finite-size effects are not, we have repeatedly achieved energies
which are in reasonable agreement with their experimental
counterparts, with differences attributable to vibrational
corrections. We have investigated using
DFT to relax excited-state geometries. It too is important, having the
largest impact in the H$_2$ ($\sim 0.8$ eV) and O$_2$ ($\sim 0.5$ eV)
dimers, of the molecules we have studied.  For the parahydrogen
molecule we performed DMC calculations of the ionization potential
with the protons treated as distinguishable quantum particles, finding
excellent agreement with experiment.  This demonstrates the
fundamental importance of geometrical and vibrational effects when
comparing \textit{ab initio} gaps with experiment.

We have probed the effects of fixed-node errors in SJ-DMC energy-gap
calculations for atoms, molecules, and solids, finding that the
inclusion of backflow functions generally improves DMC energy gaps in
these systems (especially in solids, where backflow lowers gaps by
0.1--0.2 eV)\@.  We have shown that, in the case of Si, the use of
backflow functions reoptimized in anionic and cationic states is
crucial in order to achieve reasonable agreement with experiment.
Residual overestimates (of order 0.5 eV for first-row atoms) are
expected in solids due to the presence of vibrational effects, which
are the dominant remaining source of uncertainty when it comes to
comparison with experiment.  We have also performed gap calculations
for free-standing monolayer phosphorene, showing that systematic
finite-size effects are qualitatively different in 2D materials.

\begin{acknowledgments}
R.J.H.\ is fully funded by the Graphene NOWNANO CDT (EPSRC Grant No.\
EP/L01548X/1). M.S.\ was funded by the EPSRC standard grant ``Non-perturbative
and stochastic approaches to many-body localization'' (EPSRC Grant No.\
EP/P010180/1). Computer time was provided by Lancaster University's High-End
Computing facility, the N8 HPC (funded by the N8 consortium and the EPSRC,
Grant No.\ EP/K000225/1), the ARCHER UK National Supercomputing Service, and by
the facilities of the Center for Information
Science at JAIST\@. R.M.\ is grateful for financial support from MEXT-KAKENHI
(17H05478 and 16KK0097), from FLAGSHIP2020 (Project Nos.\ hp180206 and hp180175
at K-computer), from Toyota Motor Corporation, from I-O DATA Foundation, and
from the Air Force Office of Scientific Research
(AFOSR-AOARD/FA2386-17-1-4049). We acknowledge useful discussions with Sam
Murphy, Matthew Foulkes, and Richard Needs.
\end{acknowledgments}

%

\end{document}